\def\y{\mathbf{y}}

\def\z{\mathbf{z}}
\def\e{\mathbf{e}}
\def\r{\mathbf{r}}
\def\x{\mathbf{x}}
\def\g{\mathbf{g}}

\def\e{\mathbf{e}}

\def\A{\mathbf{A}}

\def\R{\mathbb{R}}
\def\bR{\mathbf{R}}
\def\t{\mathbf{t}}
\def\I{\mathbf{I}}

\documentclass[10pt,journal,compsoc]{IEEEtran}
\usepackage{amsfonts}
\usepackage{amsmath}
\usepackage[caption=false]{subfig}
\usepackage{cuted}
\usepackage{booktabs}
\usepackage{color}
\usepackage{marginnote}
\usepackage{fancyhdr}

\ifCLASSOPTIONcompsoc
  \usepackage[nocompress]{cite}
\else
  \usepackage{cite}
\fi
\ifCLASSINFOpdf
   \usepackage[pdftex]{graphicx}
\else
\fi
\usepackage{algorithm}
\usepackage{algorithmic}
\begin{document}
%
\title{Block stochastic gradient descent for large-scale tomographic reconstruction in a parallel network}
%
%
%
%

\author{Yushan~Gao,
        Ander~Biguri,
        and~Thomas~Blumensath
\IEEEcompsocitemizethanks{\IEEEcompsocthanksitem Y. Gao, A. Biguri and T. Blumensath are with the Faculty of Engineering and Environment, University of Southampton, Southampton,
UK, SO17 1BJ.\protect\\
E-mail:  y.gao, a.biguri, Thomas.Blumensath@soton.ac.uk}
\thanks{Manuscript received March 27, 2019.}}

\IEEEtitleabstractindextext{%
\begin{abstract}
Iterative algorithms have many advantages for linear tomographic image reconstruction when compared to back-projection based methods. However, iterative methods tend to have significantly higher computational complexity. To overcome this, parallel processing schemes that can utilise several computing nodes are desirable. Popular methods here are row action methods, which update the entire image simultaneously and column action methods, which require access to all measurements at each node. In large scale tomographic reconstruction with limited storage capacity of each node, data communication overheads between nodes becomes a significant performance limiting factor. To reduce this overhead, we proposed a row action method BSGD. The method is based on the stochastic gradient descent method but it does not update the entire image at each iteration, which reduces between node communication. To further increase convergence speeds, an importance sampling strategy is proposed. We compare BSGD to other existing stochastic methods and show its effectiveness and efficiency. Other properties of BSGD are also explored, including its ability to incorporate total variation (TV) regularization and automatic parameter tuning.     
\end{abstract}

\begin{IEEEkeywords}
CT image reconstruction, parallel computing, gradient descent, coordinate descent, linear inverse problems.
\end{IEEEkeywords}}

\maketitle

\IEEEdisplaynontitleabstractindextext

%
\IEEEpeerreviewmaketitle

\section{Introduction}\label{sec:introduction}
\IEEEPARstart{I}{n} transmission X-ray computed tomography (CT), when using non-standard scan trajectories or when operating with high noise levels, traditional analytical reconstruction techniques such as the filtered backprojection algorithm (FBP) \cite{sagara2010abdominal,hoffman1979quantitation} and the Feldkamp Davis Kress (FDK) \cite{feldkamp1984practical,rodet2004cone} method are no longer applicable. In these circumstances, less efficient, iterative reconstruction methods can provide significantly better reconstructions \cite{gervaise2012ct,wang2008outlook,deng2009parallelism,
willemink2013iterative}. These methods model the x-ray system as a linear system:
\begin{equation}
\y=\A \x+\mathbf{e},
\label{yax}
\end{equation}
where $\y=[y_1,\cdots,y_r]^T,\x=[x_1,\cdots,x_c]^T$ and $\e=[e_1,\cdots,e_r]^T$ are x-ray projection data, the unknown vectorised image and measurement noise respectively. The system matrix $\A\in \mathbb{R}^{r*c}$ has non-negative elements, which can be computed using Siddon's method \cite{jacobs1998fast}. Image reconstruction can then be cast as an optimisation problem \cite{soleimani2015introduction,guo2016convergence,beister2012iterative}:
\begin{equation}
\min_\x f(\x)=\min_\x \frac{1}{2}\|\y-\A\x\|_2^2,
\label{min}
\end{equation}

 In many applications, such as industrial CT scanning, the system matrix $\A$ can be enormous \cite{ni2006review}.  Iterative methods apply matrices $\A$ and $\A^T$ to compute ``forward projection''(FP) and ``back projection''(BP) respectively, to iteratively find an approximate solution to minimize Eq.\ref{min}. Note that in realistic applications, due to its size, the matrix $\A$ is never stored \cite{van2015astra}, FP and BP are instead computed `on the fly' using Graphical Processor Units (GPUs).

 For our discussion, we classify iterative methods into column action methods  and row action methods. 
Column action methods include iterative coordinate descent (ICD)\cite{yu2011fast,benson2010block} and axial block coordinate descent (ABCD) \cite{fessler2011axial,kim2012parallelizable}. They divide $\x$ into several blocks and update individual blocks in each iteration using the most recent estimates of all other blocks. Row action methods include Kaczmarz methods(ART) \cite{li2013adaptive}, simultaneous iterative reconstruction technique(SIRT) \cite{gregor2008computational}, and component averaging (CAV) \cite{censor2001component} and their ordered set variations \cite{censor2001bicav,xu2010efficiency}.   Unlike column action methods, row action methods divide the projection data $\y$ into several blocks and update all of $\x$ simultaneously using one or several blocks of $\y$.  
Despite the superior reconstructions achievable with iterative methods in many applications, the high computational complexity remains a significant bottleneck limiting their application in realistic settings. To overcome these issues, parallization is desirable. For example, the ICD algorithm can be run on multiple CPUs \cite{wang2016high} or on several graphics processing units(GPUs) \cite{sabne2017model,wang2017massively}. Parallization of row action methods  is straightforward: each node receives a copy of $\x$ and different blocks of $\y$. Each node independently updates $\x$ and message passing between nodes computes weighted sums of partial results \cite{bilbao2004performance,flores2012fast}.  Compared with column action methods, row action methods is more amenable to parallel processing in a multiple-node network because that  different nodes do not have to update the same elements in the error vector $\y-\A\x$\cite{rui2012evaluation}. 

A range of row and column action methods have been specifically designed for tomographic reconstruction. Recently, advances in machine learning have also led to significant advances in stochastic optimization and many of these ideas are also applicable to tomographic reconstruction. Most methods here are row action methods. These include stochastic gradient descent \cite{ruder2016overview}, stochastic variance reduced gradient(SVRG) \cite{johnson2013accelerating}, incremental aggregated gradient (IAG) \cite{IAG} and stochastic average gradient (SAG)\cite{SAG}. These stochastic algorithms have often been parallelized to operate on large data sets \cite{recht2011hogwild,zhao2016fast,leblond2016asaga,zhang2015fast}.

\section{The BSGD algorithm}
In this paper, we develop a parallel row action algorithm specifically for large scale tomographic reconstruction. Whilst previous work in parallel tomographic reconstruction has concentrated on standard tomography, where an object is rotated around a single axis, we are here particularly interested in a setting that allows more general trajectories such as those found in laminographic scanning \cite{Woodetal2018}. With ``large scale'' we here mean that both $\y$ and $\x$ are too large to be stored within one computing node. We thus divide $\y$, $\A$ and $\x$ into several blocks and design algorithms in which each node only has partial access to both $\y$ and $\x$ at each iteration. Let $\A$ be divided into $M$ row blocks and $N$ column blocks (possibly after row and column permutation). Let $\{I_i\}_{i=1}^M$ be a set of row indices and $\{J_j\}_{j=1}^N$ a set of column indices.  $\A_I^J$ thus is a sub-matrix  indexed by $I\in \{I_i\}_{i=1}^M$ and $J\in \{J_j\}_{j=1}^N$. The forward X-ray projection process can then be approximated as:  
\begin{equation}
\begin{bmatrix}
\y_{I_1}\\
\vdots\\
\y_{I_M} 
\end{bmatrix}\approx
\begin{bmatrix}
\A_{I_1}^{J_1} & \cdots & \A_{I_1}^{J_N}\\
\vdots & \vdots & \vdots \\
\A_{I_M}^{J_1} & \cdots & \A_{I_M}^{J_N}
\end{bmatrix}
\begin{bmatrix}
\x_{J_1}\\
\vdots \\
\x_{J_N}
\end{bmatrix}\equiv \begin{bmatrix}
\A_{I_1}\\
\vdots \\
\A_{I_M}
\end{bmatrix}\x.
\label{e1}
\end{equation}

With this partitioning, both column and row action methods can be inefficient in terms of communication between nodes since they all require full access to either all of $\y$ or all of $\x$. For example, in row action methods, the most time consuming operations are the FP and BP \cite{palenstijn2015distributed}. These projections are parallelizable \cite{rosen2013iterative}. For the FP $\A_I\x\equiv \sum_{j=1}^N \A_I^{J_j}\x_{J_j}$, each parallel node calculates a forward projection $\A_I^{J_j}\x_{J_j}$. The summation over $j$ is then calculated at a master node or using an ALLREDUCE  procedure \cite{jones2006hybrid}. A similar parallel scheme is also applicable to the BP. If the number of computing nodes is smaller than the number of column or row blocks $N$ or $M$, then the parallel calculations require several communications between data storage and computation nodes, which can be time consuming \cite{bilbao2004performance,deng2009parallelism}. 

To reduce the communication overhead and the algorithm's dependency on  access to all of $\y$ or $\x$, we previously proposed a parallel algorithm called Coordinate-Reduced Steepest Gradient Descent (CSGD) \cite{gao2018joint} to solve the block linear model Eq.\ref{e1}. However, our previous method only converged to a weighted least squares solution. Here we propose an improved algorithm, which we call `Block Stochastic Gradient Descent'' (BSGD). We empirically show that BSGD converges closer to the least squares solution than CSGD. The new method is similar to SAG and accumulates previously calculated direction to obtain the current update direction, but it differs from SAG in that we also incorporate a coordinate descent strategy. In the origin SAG algorithm, the gradient summands must be calculated by accessing all of $\x$ while in BSGD only part of $\x$ is required and updated. Furthermore, we exploit the sparsity of the system matrix $\A$ found in CT imaging and proposed an ``importance sampling'' strategy for BSGD (BSGD-IM). An automatic parameter tuning strategy is also adopted to tune the step length. Simulation results show that the convergence speed of BSGD-IM is faster than other row action methods such as SAG and SVRG, making BSGD an ideal candidate for distributed tomographic reconstruction.

We derive BSGD for large scale CT reconstruction where a parallel multiple-GPU network is available. The network uses a master-servant architecture and the servants/nodes(i.e. GPUs) in the network have limited access to both projection data $\y$ and reconstructed volume $\x$. 
To facilitate latter discussions, we define $\r=\y-\A\x$ and let $\r_I$ be the subset of $\r$ representing $\y_I-\A_I\x$.

\begin{figure}[htb]
\centering     
\includegraphics[width=55mm]{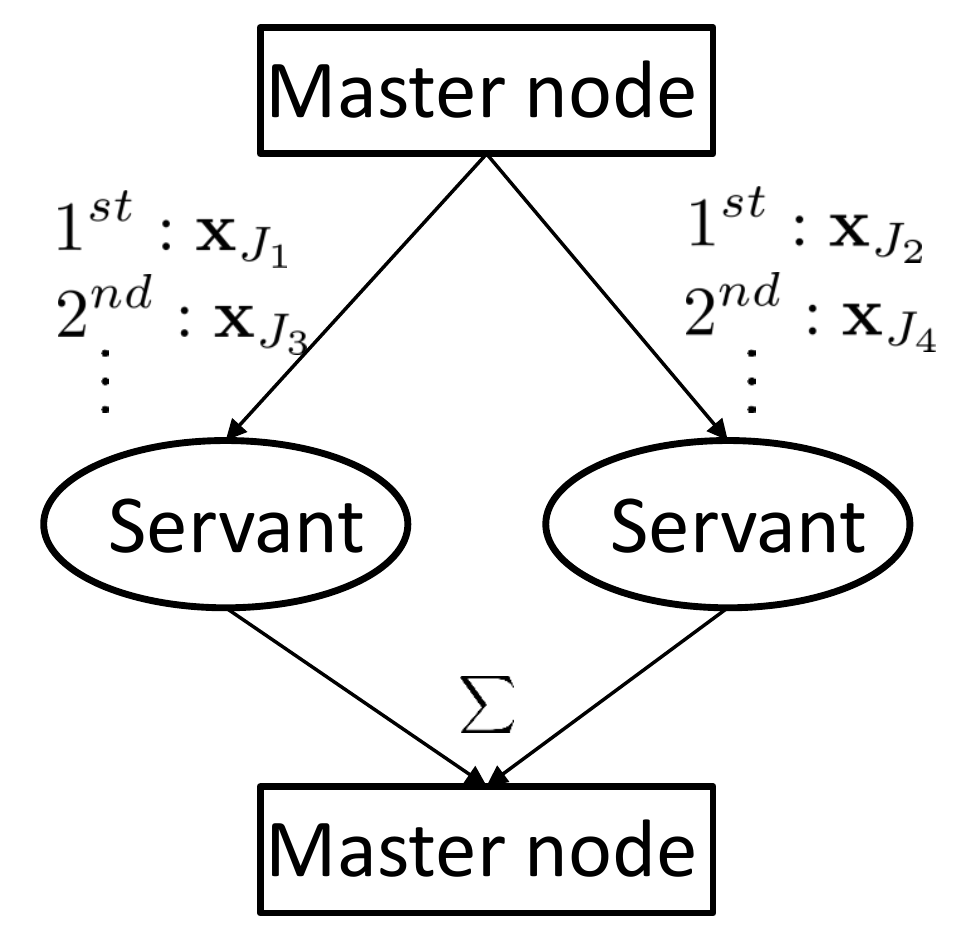}
\caption{In a master-servant network, in each iteration, each servant node receives one block $\x_{J}$ to calculate $\A_I^J\x_J$. The results are accumulated in the master node to update the corresponding residual $\r_I$. If the number of the servant nodes is less than the number of column blocks $N$, then we require repeated communication between servants and the master nodes.}
\label{FigParalProtype}
\end{figure}

To motivate our approach, let us consider iterative mini-batch stochastic gradient descent \cite{li2014efficient} with a decreasing step length $\mu$. At the $k^{th}$ iteration, this method computes:
\begin{equation}
\begin{aligned}
& \r_I=\y_I-\A_I\x \\
& \g = 2\A_I^T\r_I \\
& \x^{k+1}=\x^k+\mu\g.\\
\label{SGDITE}
\end{aligned}
\end{equation}
In our setting where a master or storage node assigns data to servant nodes for processing, if local memory at a servant node is restricted, then each node can only process a partial block $\x_J$ and calculate $\A_I^J\x_J$. To compute $\A_I\x$, we would thus need to repeatedly sent different blocks to the servant nodes before the result is combined in the master node to update $\r_I$. This process is shown in Fig.\ref{FigParalProtype}.
Computation of $\A_I^T\r_I$ would need to follow a similar strategy. 
Instead of updating $\x$ only once the exact residual $\r$ has been computed, our innovation is to compute a stochastic approximation of the residual by only processing a subset of $\x$ in each iteration and using previously computed estimates of $\A_I^J\x_J$ for the blocks not used in this step. The hope is that the increase in uncertainty in the gradient estimate is compensated for by a reduction in computation and communication cost.
 BSGD thus does not compute all $\A_I^T\r_I$ and $\A_I\x_I$ in each iteration. For a fixed number of servant nodes, let $\alpha$ and $\gamma$ be the fraction of row and column blocks that can be used in parallel computations at any one time. During each iteration, we thus propose to only use $\alpha \gamma MN$ sub-matrices ($\A_I^J$) together with the corresponding data ($\y_I$) and volume($\x_J$)  sub-vectors to compute updates. 
 To gradually reduce the error between the stochastic gradient and the true gradient, BSGD adopts a gradient aggregation strategy that is similar to that of  SAG. As we show below for our CT reconstruction problem, this accumulation strategy enables the algorithm to converge with a constant step length $\mu$. 
 
 The BSGD algorithm is shown in Algo.\ref{alg1}. When $\gamma=1$, the method becomes SAG. In this paper we mainly focus on $\alpha$ and $\gamma<1$, which allows us to use a reduced number of servant nodes.
\begin{algorithm}[htb]  
  \caption{BSGD}
   \label{alg1}  
    \begin{algorithmic}[1]
    \STATE Initial: $\g=\mathbf{0},\{\hat{\g}^i\}_{i=1}^M=\{\mathbf{0}\},\{\z^j\}_{j=1}^N=\{\mathbf{0}\}, \r=\y$, $\mu=$ const. $\x_{est}=\mathbf{0}$.
    \FOR{epoch =1,2,..., Max Iteration}
    	\STATE randomly select $\alpha M$ row blocks from $\{I_i\}_{i=1}^M$ and $\gamma N$ column blocks from $\{J_j\}_{j=1}^N$
  		\FOR{the selected $I_i$ and $J_j$ \textbf{in parallel}}
			\STATE $\z_{I_i}^j=\A_{I_i}^{J_j}{\x_{est}}_{J_j}$ 
  		\ENDFOR
        \STATE $\r=\y-\sum_{j=1}^N\z^j$
        \FOR{the selected  $I_i$ and $J_j$ \textbf{in parallel}}
        	\STATE ${\hat{\g}}_{J_j}^i=2(\A_{I_i}^{J_j})^T\r_{I_i}$ 
        \ENDFOR
        \STATE $\g=\sum_{i=1}^M\hat{\g}^i$
        \FOR {the selected $J_j$ \textbf{in parallel}}
        \STATE ${\x_{est}}_{J_j}={\x_{est}}_{J_j} + \mu\g_{J_j}$
        \ENDFOR
    \ENDFOR
\end{algorithmic}  
\end{algorithm}
\subsection{Improving BSGD performance}
\begin{figure}[htb]
\centering     
\includegraphics[width=60mm]{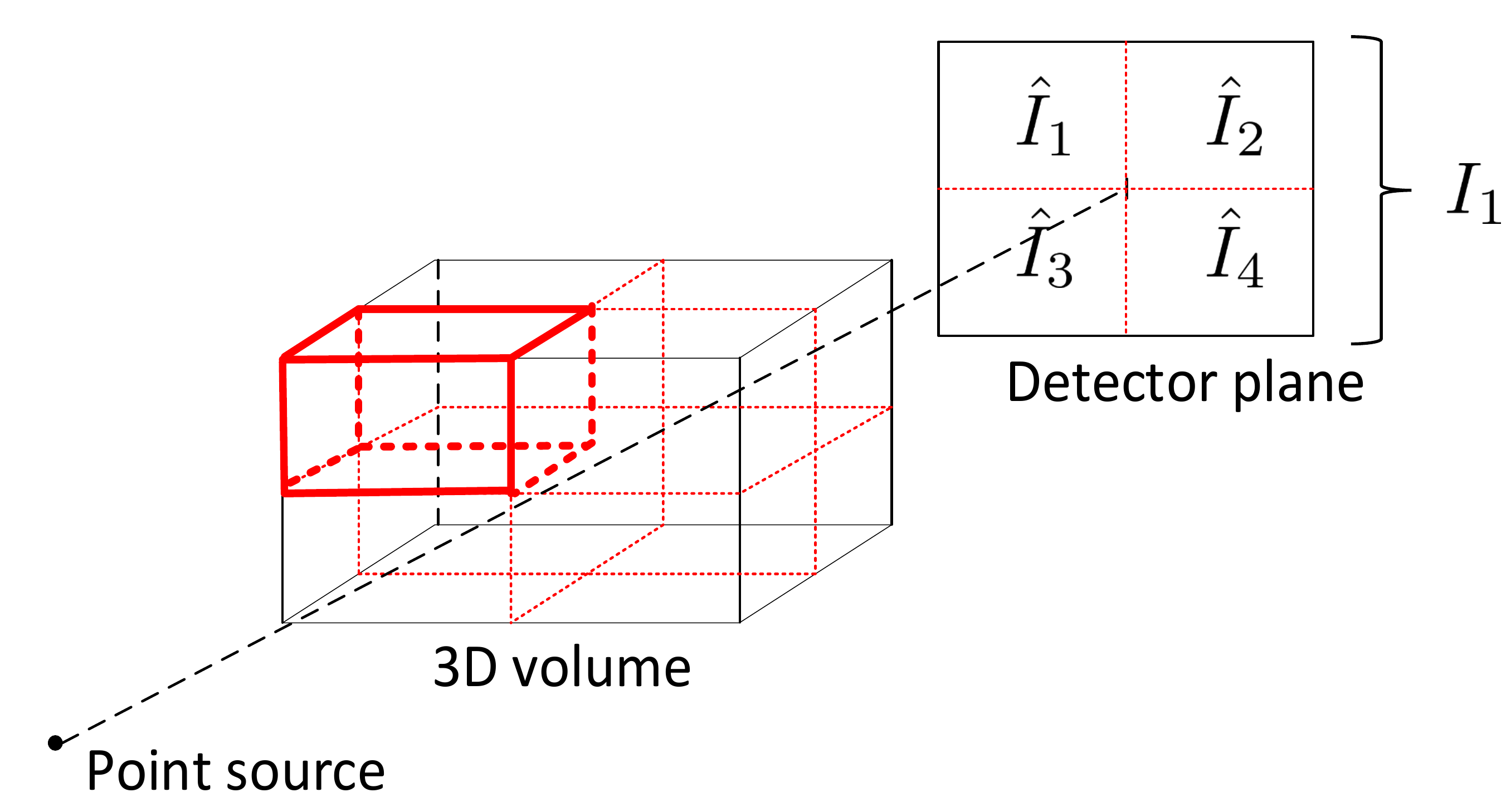}
\caption{Example cone beam CT setting. The 3D volume is divided into 8 sub volumes. In the cone beam scanning geometry, the projection of one sub-block is mainly concentrated in a small area on the detector. If the detector is also divided into 4 sub-areas,  then the currently selected sub-volume (bold red frame) is mainly projected onto the top-left and top-right sub-detector area, i.e. $\hat{I}_1$ and $\hat{I}_2$ area.}
\label{Impor}
\end{figure}

There are two tricks to improve BSGD performance. The first trick is the use of an importance sampling strategy which we call ``BSGD-IM''. In tomographic reconstruction, the matrix $\A$ is often sparse with different blocks $\A_I^J$ often varying widely in their sparsity. 
This sparsity can be enhanced further for 3D tomographic problems if we partition $\y$ such that each partition is made up of a selection of sub-blocks, where each sub-block corresponds to a partition of the X-ray detector at one projection angle as shown in Fig. \ref{Impor}.
Inspired by this property, BSGD-IM breaks each single projection into several sub-projections (4 sub-projections in Fig.\ref{Impor}) and only samples 1 sub-projections for the selected sub-volume $\x_J$. This sampling is not done uniformly but is based on a selection criteria that uses the relative sparsity of each matrix $\{\A_{\hat{I}_i}^J\}$, i.e. the denser a sub-matrix is, the higher the probability that it is selected. 
As we do not have access to matrix $\A$, to estimate the sparsity pattern of $\A$, BSGD-IM computes the fraction of a volume block's projection area on each sub-detector. When the row block $I_i$ contains several projection angles, the importance sampling strategy is repeatedly applied to each projection angle contained in the row block. The advantage of BSGD-IM is that it provides each computation node more projection angles within one row block than BSGD when a nodes' storage capacity is limited and thus reduces the row block number $M$ as well as the total storage requirement. For example, if we assume that one GPU can only process a single projection in the original BSGD, then BSGD-IM with the partition of Fig.\ref{Impor} can use four projection angles by choosing one detector sub-areas for each projection angle. 

BSGD-IM is shown in Algo.\ref{alg2}. Whilst importance sampling speeds up initial convergence, it also introduces a bias is the stochastic gradient due to the inhomogeneous sampling. To overcome this, it is suggested that after initial fast convergence, the last few iterations should be run without importance sampling.

\begin{algorithm}[htb]  
  \caption{BSGD-IM}
   \label{alg2}  
    \begin{algorithmic}[1]
     \STATE Initial: $\g=\mathbf{0},\{\hat{\g}^i\}_{i=1}^M=\{\mathbf{0}\},\{\z^j\}_{j=1}^N=\{\mathbf{0}\}, \r=\y$, $\mu=$ const. $\x_{est}=\mathbf{0}$. 
     \FOR{epoch =1,2,..., Max Iteration}
    	\STATE randomly select $\alpha M$ row blocks from $\{I_i\}_{i=1}^M$ and  $\gamma N$ column blocks from $\{J_j\}_{j=1}^N$.
  		\FOR{the selected $I_i$ and $J_j$ \textbf{in parallel}}
  		\STATE For each column block $\x_{J_j}$, importance sample one sub-detector area from each single projection view. All indexes represented by those selected sub-detector areas form the row index set $\hat{I}_t$.
			\STATE $\z_{\hat{I}_t}^j=\A_{\hat{I}_t}^{J_j}{\x_{est}}_{J_j}$
  		\ENDFOR
        \STATE $\r=\y-\sum_{j=1}^N\z^j$
        \FOR{the selected $I_i$ and $J_j$ \textbf{in parallel}}
        	\STATE ${\hat{\g}}_{J_j}^i=2(\A_{\hat{I}_t}^{J_j})^T\r_{\hat{I}_t}$ 
        \ENDFOR
        \STATE $\g=\sum_{i=1}^M\hat{\g}^i$
        \FOR {the selected $J_j$ \textbf{in parallel}}
        \STATE ${\x_{est}}_{J_j}={\x_{est}}_{J_j} + \mu\g_{J_j}$
        \ENDFOR
    \ENDFOR
\end{algorithmic}  
\end{algorithm}

The second trick is to use automatic parameter tuning. 
Broadly speaking, up to a limit, increasing $\mu$ increases convergence speed. However, in practice, it is difficult to determine the upper limit. As a result, in realistic large scale tomographic reconstruction, instead of using a fixed step-length  $\mu$, we  developed an automatic parameter tuning approach.    
Parameter tuning is not a new concept in machine learning and optimization. For example, the hypergradient descent\cite{baydin2017online} or the Barzilai-Borwein (BB) method \cite{tan2016barzilai} can be used for SGD or SVRG. However these methods are not directly applicable to BSGD, as they require updates to all of $\x$ in each iteration. Furthermore, BSGD uses dummy variables $\z$ to stores information about previous $\x$.  Due to this, the stochastic gradient $\g$ of BSGD is much noisier than the estimate obtained by traditional stochastic gradient methods. 
We here proposed an automatic parameter tuning methods that is different from the BB method or hypergradient descent method. It only exploits the parameters generated during the iteration process: the residual $\r$ and iteration direction $\g$, as shown in Algo.\ref{cre1}.
\begin{algorithm}[htb]  
  \caption{Automatic $\mu$ tunning strategy}
  \label{cre1}  
    \begin{algorithmic}[1]
    \STATE	$\epsilon$ and $\delta$ are positive constants.
    \STATE At each $k^{th}$iteration where $mod(k,M)==0$, sum up all $\g$ in the past $M$ epochs as an effective update direction (EUD) $\overline{\g}^{k/M}$. 
    \STATE calculate the inner product between two consecutive $\overline{\g}$ as $\theta^{k/M}=\frac{(\overline{\g}^{k/M})^T\overline{\g}^{k/M-1}}{\|\overline{\g}^{k/M}\|\|\overline{\g}^{k/M-1}\|}$.
    \IF {mod($k$,$M$)==$0 \& k > M$}
        \IF{$\|\r\|^{k}<\|\r\|^{k-M}<\|\r\|^{k-2M}$}
	    \STATE $\mu=(1+\epsilon)*\mu$
		\ENDIF
		\IF{$\|\r\|^{k}>\|\r\|^{k-M}>\|\r\|^{k-2M}$ }
			\IF {$\mid\theta^{k/M}-\theta^{k/M-1}\mid > t_1$ or $\theta^{k/M}<t_2$}
		    \STATE $\mu=(1-\delta)\mu$
		    \ENDIF
		\ENDIF
    \ENDIF
\end{algorithmic}  
\end{algorithm}

This automatic parameter tuning is applied after $M$ iterations. It tests whether $\r$ decreased during the past $2M$ epochs, in which case $\mu$ is increased by $1+\epsilon$. To reduce $\mu$, using a similar condition on $\r$ alone (i.e. line 8 in Algo.\ref{cre1}, named as ``criteria 1'') was not found to be sufficient to ensure convergence. We thus use an additional criteria (line 9 in Algo.\ref{cre1}, named as ``criteria 2''). The criteria 2 is motivated by the general parameter tuning methods that computes inner-products between adjacent gradients and determines to increase or decrease  $\mu$ according to the positivity of the inner-product \cite{baydin2017online,plakhov2004stochastic}.  We thus compare the inner-products of two gradients. To do this, we accumulate several stochastic gradients during a period of several iterations ($M$ epochs in Algo.\ref{cre1}) to compute an effective update direction (EUD) $\overline{\g}$ to reduce the stochastic error variance. We have observed that when BSGD converges with a properly chosen fixed $\mu$, then the change of two adjacent EUDs do not vary significantly. On the contrary, these two directions vary significantly when  BSGD suffers from oscillatory behaviour or an increase in the norm of $\r$. This is due to our method using some old values $\z$ in the calculation of each update. If the change in $\x$ is not too large, then these old values for $\z$ are good approximations to the current values. As a result, the two EUDs, should also be similar to each other. If we assume that during $M$ epochs it is likely that all of $\x$ (and thus all of the $\z$'s) have been updated, then two EUDs can be computed from the previous $2M$ epochs. The inner product of the two normalized EUDs should always be close to 1. If not, it means that the step length is too big and the iteration is likely to diverge.

For both increasing and decreasing $\mu$ part, the frequency of parameter changing is $M$ epochs rather than 1 epoch. One reason is that the high stochastic noise effect in the gradient update can be reduced after a period of epochs. Another reason is that the calculation of $\|\r\|$ can be time consuming when the size of $\r$ is large, reducing the frequency of computation on $\|\r\|$ is beneficial to save the reconstruction time. We have experimentally validated that setting the test frequency to $M$ leads to a good compromise between increased computational demand and improved overall convergence speed.
\subsection{Incorporating TV regularization into BSGD}
When we have few projections, a TV regularization term is often used to increase the reconstruction quality. Reconstruction with a TV regularization term often minimizes a quadratic objective function plus a non-smooth TV-regularization term:
\begin{equation}
\underbrace{(\y-\A\x)^T(\y-\A\x)}_{f(\x)}
+\underbrace{2\lambda \text{TV}(\x)}_{g(\x)},
\label{TVequ}
\end{equation}
where $\lambda$ is a relaxation parameter and $\text{TV}(\x)$ is the total variation (TV) of $\x$. For 2D images, the total variation penalty can be defined as: 
\begin{equation}
\text{TV}(\x)=\sum_{c,d}\sqrt{(x_{c,d}-x_{c-1,d})^2+(x_{c,d}-x_{c,d-1})^2},
\end{equation}
where $x_{c,d}$ is the intensity of image pixel in row $c$ and column $d$.

Traditional methods, including ISTA\cite{combettes2005signal}  and FISTA\cite{beck2009fast},  minimize the TV regularized objective function with two steps: Each iteration starts with using the gradient of $f(\x)$ to reduce the data fidelity, i.e. to reduce $f(\x)$, followed by a TV-based de-noising procedure. We empirically show that  BSGD is able to replace the gradient descent (GD) step. Since BSGD updates only some components of $\x$ with partial projection data in each iteration, the TV-based de-noising procedure is only performed after a period of time, enabling the computation load of BSGD is scalable to that in ISTA or FISTA. The algorithm is shown in Algo.\ref{BSGDTV}.
\begin{algorithm}[htb]  
  \caption{BSGD-TV}
   \label{BSGDTV}  
    \begin{algorithmic}[1]
    \STATE Initial: $\g=\mathbf{0},\{\hat{\g}^i\}_{i=1}^M=\{\mathbf{0}\},\{\z^j\}_{j=1}^N=\{\mathbf{0}\}, \r=\y$, $\mu=$ const. $\x_{est}=\mathbf{0}$.
    \FOR{epoch =1,2,..., Max Iteration}
    	\STATE randomly select $\alpha M$ row blocks from $\{I_i\}_{i=1}^M$ and $\gamma N$ column blocks from $\{J_j\}_{j=1}^N$
  		\FOR{the selected   $I_i$ and $J_j$ \textbf{in parallel}}
			\STATE $\z_{I_i}^j=\A_{I_i}^{J_j}{\x_{est}}_{J_j}$ 
  		\ENDFOR
        \STATE $\r=\y-\sum_{j=1}^N\z^j$
        \FOR{the selected  $I_i$ and $J_j$ \textbf{in parallel}}
        	\STATE ${\hat{\g}}_{J_j}^i=2(\A_{I_i}^{J_j})^T\r_{I_i}$ 
        \ENDFOR
        \STATE $\g=\sum_{i=1}^M\hat{\g}^i$
        \FOR {the selected $J_j$ \textbf{in parallel}}
        \STATE ${\x_{est}}_{J_j}={\x_{est}}_{J_j} + \mu\g_{J_j}$
        \ENDFOR
        \IF{$mod(k,\frac{1}{\alpha\gamma})==0$}
        \STATE $\x_{est}=\arg\min_{\t} \Vert\t-\x_{est}\Vert^2+2\mu\lambda \text{TV}(\t)$
        \ENDIF
    \ENDFOR
\end{algorithmic}  
\end{algorithm}


\section{Results}
We start the evaluation of our method using a 2D scanning setup (sections \ref{A} to \ref{D}) to explore convergence properties of BSGD. In the final section (\ref{E}), we then look at a more representative 3D cone beam setting. 
\subsection{BSGD convergence}
\label{A}
\begin{figure}[htb]
\centering     
\includegraphics[width=60mm]{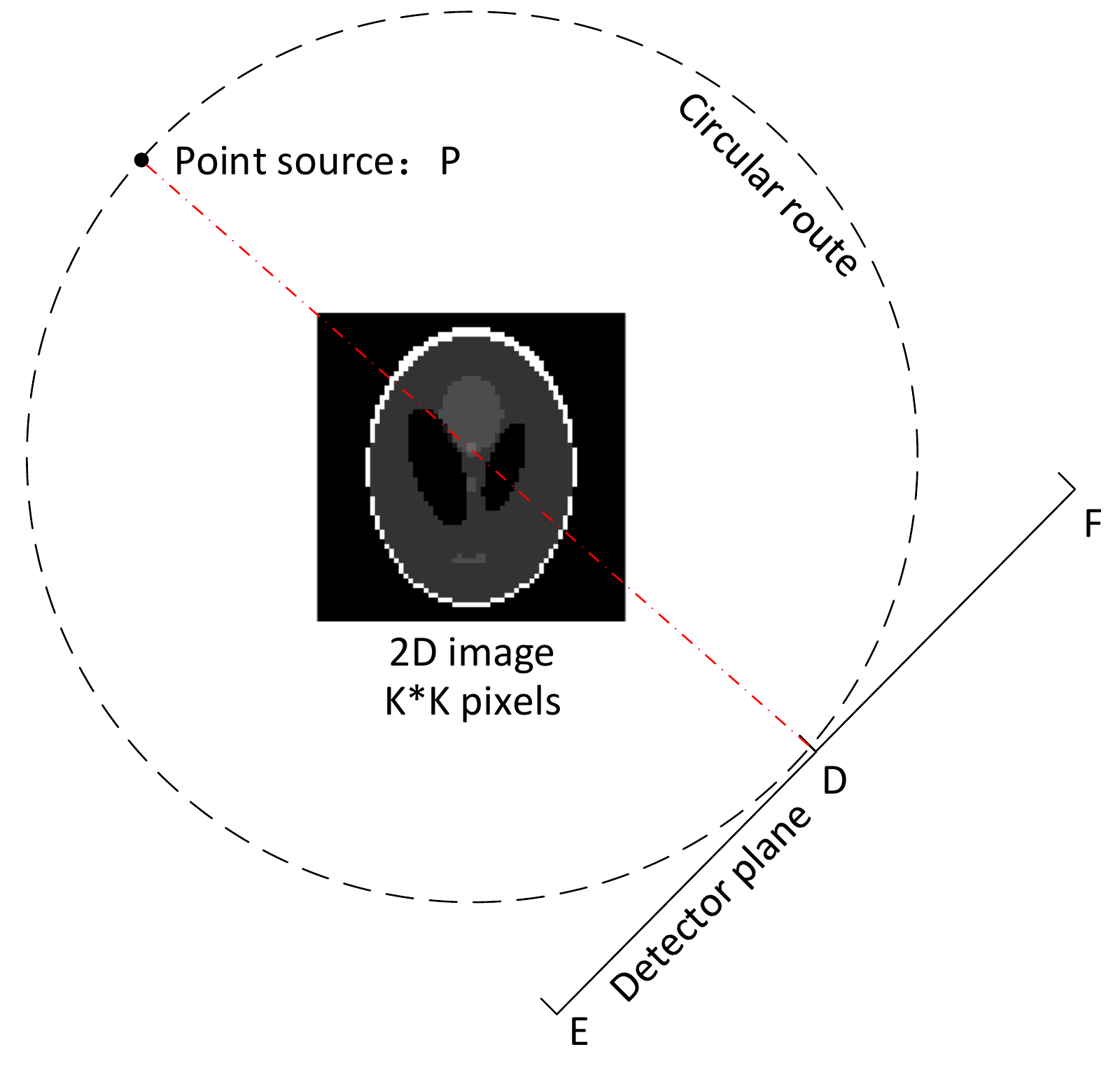}
\caption{A standard 2D scanning geometry with a Shepp-Logan phantom, where P is the x-ray source, O is the centre of the object and the rotation centre. D is the centre of the detector. Source and detector rotate around the centre and take measurements at different angles. The linear detector is evenly divided into to sub-areas DE and DF, which will be used in importance sampling discussed later. In this paper, unless particularly mentioned, the size of the image pixels (or voxels in 3D) and the detector pixel size are both 1.
}
\label{Scanning geometry}
\end{figure}

We here use the scanning geometry as shown in Fig.\ref{Scanning geometry}. 
In our first  experiments we set $K$ to 16, $OP$ and $OD$ is 50, the detector has 30 elements and the angular interval is $10^\circ$ so that $\A\in\mathbb{R}^{1080*256}$. This model is used from section \ref{A} to section \ref{AutParaTun}.

We first examine  if  BSGD (without additional regularisation) converges to the least square solution of the linear model.
We here set $M=4$ and $N=2$. $\alpha$ and $\gamma$ are initially set to 1. We add Gaussian noise to the data so that the Signal to Noise ratio (SNR) of $\y$ is 17.5 dB. The  distance to the least square solution (\textbf{DS}) is defined as 
\begin{equation}
DS=\Vert \x_{rec}-\x_{lsq} \Vert,
\end{equation}
where $\x_{rec}$ is the reconstructed image vector and $\x_{lsq}$ is the least square solution obtained here using the LSQR method. 

\begin{figure}[htb]
\centering     
\includegraphics[width=70mm]{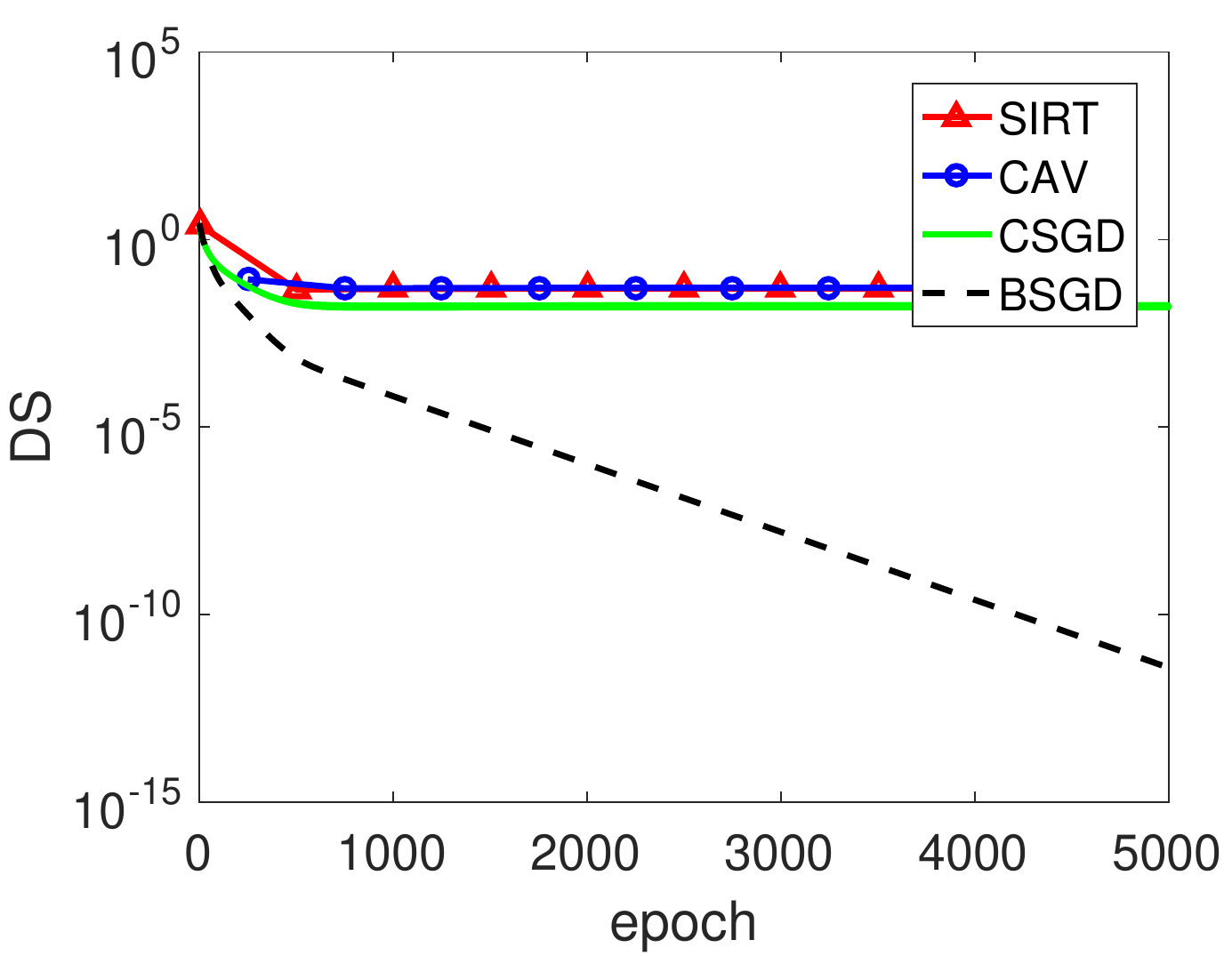}
\caption{In contrast to BSGD, SIRT, CAV and CSGD do not achieve the least square solution. }
\label{ExamSimu38}
\end{figure}

We compared BSGD with other mature methods including SIRT and CAV as well as with our previous algorithm CSGD. The results are shown in Fig.\ref{ExamSimu38}, where we see the linear convergence of our method to the least squares solution. 
All parameters in the methods were well tuned to ensure the fastest convergence rate. We see that BSGD not only approaches the least square solution, it also shows a faster convergence rate compared to the other methods.

\begin{figure}[htb]
\centering     
\includegraphics[width=70mm]{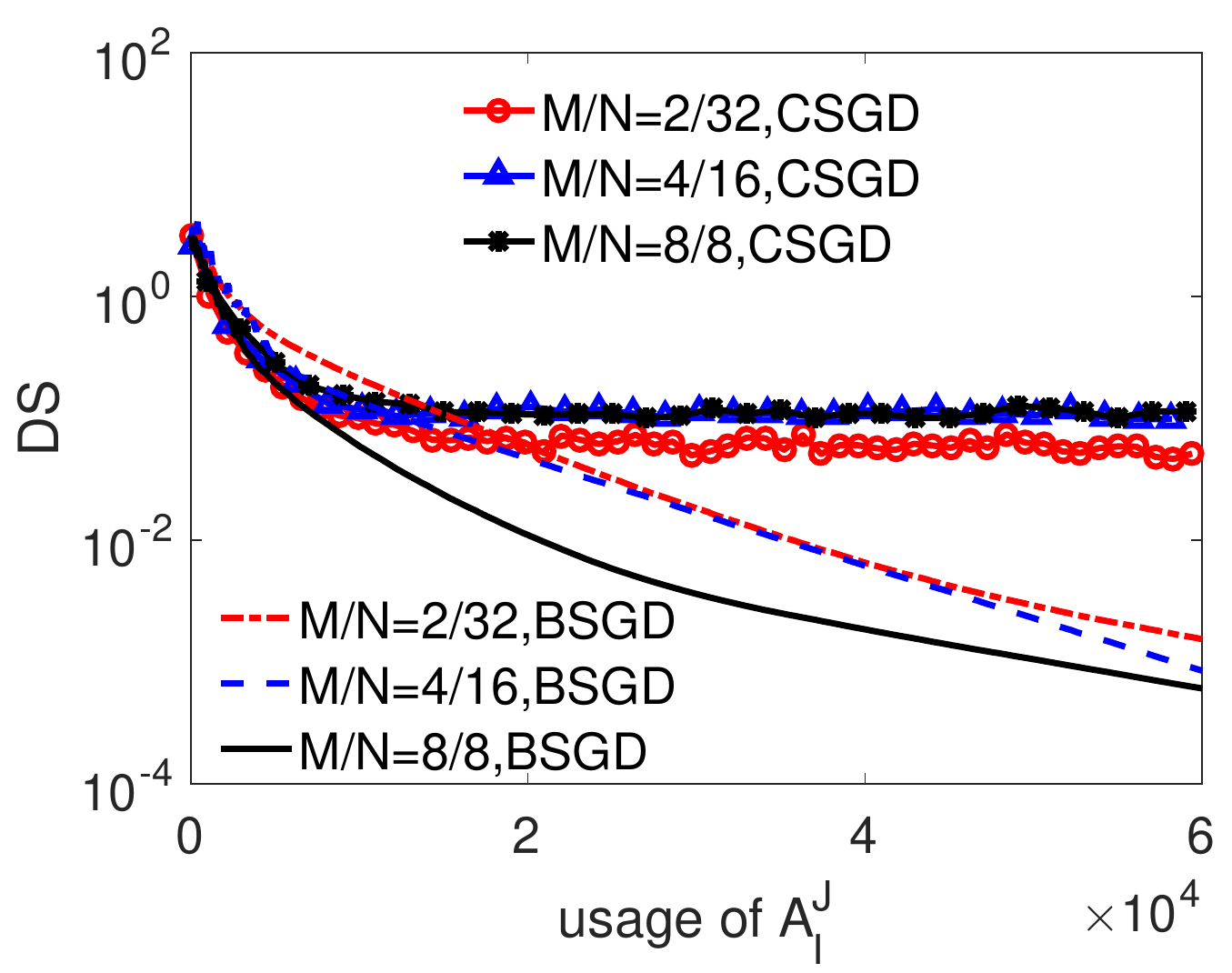}
\caption{BSGD shows better converge compared to CSGD in terms of achieving the least squares solution. Different ways to partition the rows and columns lead to different convergence speeds.}
\label{Algo2toLsqb}
\end{figure}

The initial simulations used $\alpha = \gamma =1$. We thus next study the more realistic setting where $\alpha$ and $\gamma$ are smaller than 1.The  results, shown in Fig.\ref{Algo2toLsqb}, show that even in this scenario, BSGD still approaches the least square solution.
We here divided $\A$ into 64 blocks, using different partitions that varied in the numbers of row and column blocks (2 row blocks and 31 columns blocks, 4 row blocks and 16 column blocks and 8 row blocks and 8 column blocks). We set $\alpha$ and $\gamma$ to $\frac{1}{M}$ and $\frac{1}{N}$ respectively. To measure convergence speed and communication costs between master node and computation node in a realistic parallel network, we plot the $DS$ as a function of the number of multiplications of a vector by $\A_I^J$(or $(\A_I^J)^T$), which is proportional to the number of forward/backwards projections as well as the corresponding communication time. We see from Fig.\ref{Algo2toLsqb} that BSGD performs better in terms of reaching the least squares solution. We also see differences in the convergence speed depending on the way in which we partition the matrix.
\subsection{Setting $\alpha,\gamma,M$ and $N$}
\begin{figure}[htp]
\center
\subfloat[]{%

  \includegraphics[width=70mm]{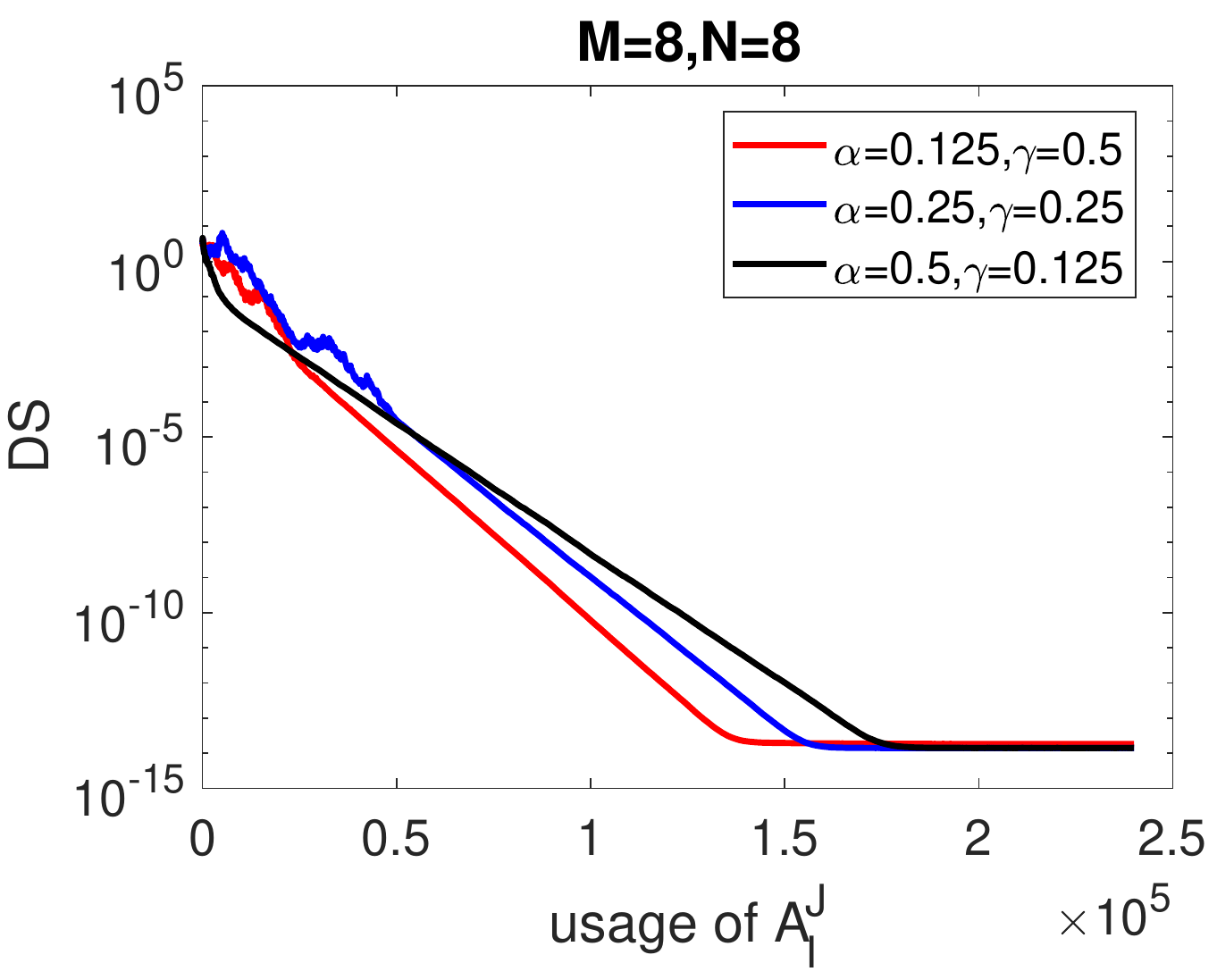}%
}

\subfloat[]{%
  \includegraphics[width=70mm]{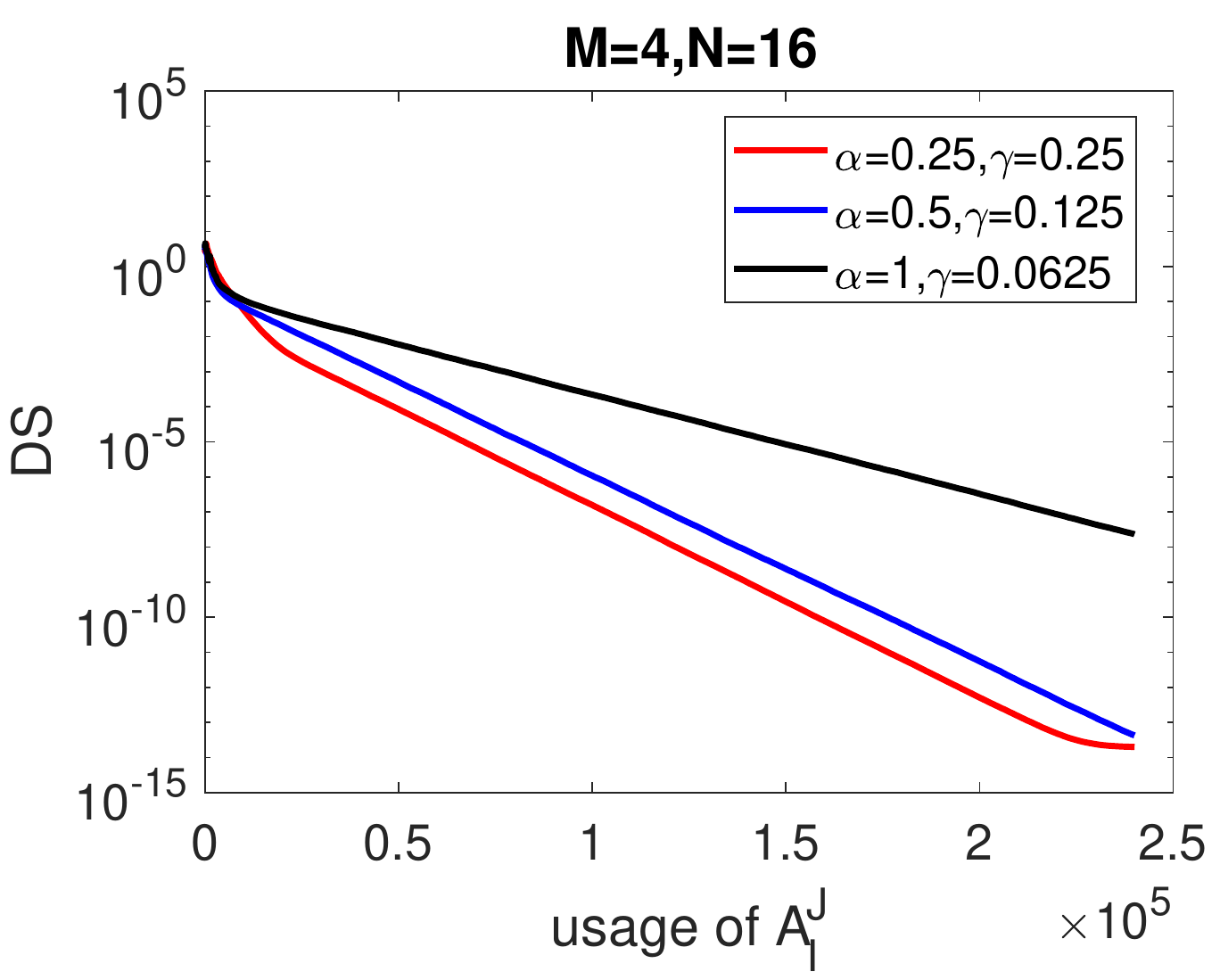}%
}
\caption{When $M$ and $N$ are fixed, reducing  $\gamma$ slows down the convergence speed.}
\label{fig1}
\end{figure}

We nexrt study the influence of $\alpha$ and $\gamma$ for a fixed partition of $\A$.  In Fig.\ref{Algo2toLsqb}  $\alpha$ and $\gamma$ were set to an extreme value where only one node is used. In realistic applications, several nodes might be available. Assume we have 4 nodes so that $MN\alpha\gamma =1$. Simulations, shown in Fig.\ref{fig1}, show that reducing $\gamma$ slows down convergence. 

Based on these and similar results, we suggest to use the following selection criteria for $\alpha$,$\gamma$
\begin{equation}
\left\{
             \begin{array}{lr}
             \gamma=\min\{1,\frac{NodeNum}{N}\},\\
             \alpha=\frac{NodeNum}{MN\gamma}, &  \\
             \end{array}
\right.
\label{Eq1}
\end{equation}
where the $NodeNum$ is the number of separate computation nodes. 

\begin{figure}[htb]
\centering     
\includegraphics[width=70mm]{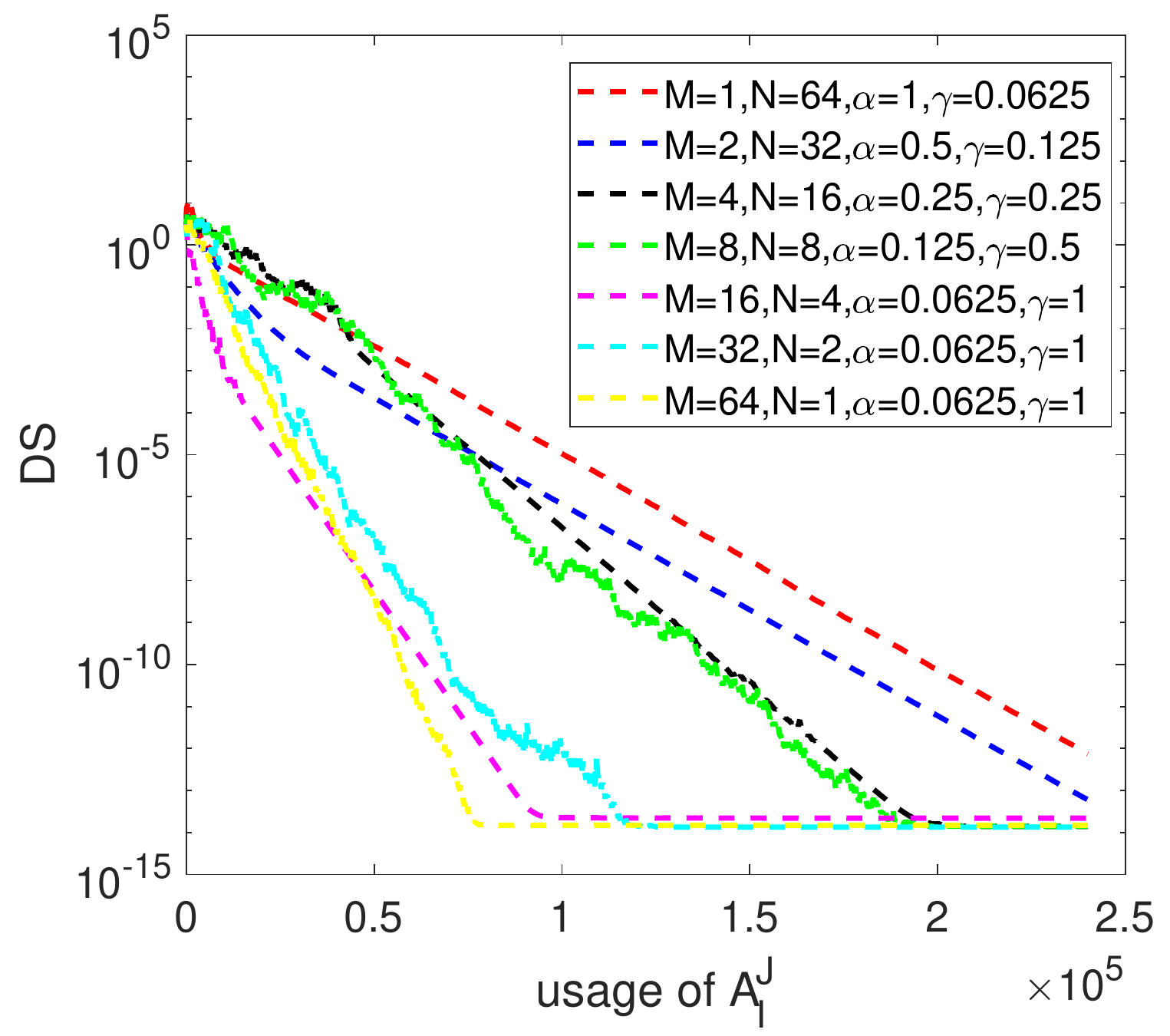}
\caption{ BSGD does not encourage the partition in the column direction. When $N$ increases, convergence speed decreases.}
\label{CompaDiffMN}
\end{figure}

\begin{figure}[htp]
\center
\subfloat[]{%
  \includegraphics[clip,width=70mm]{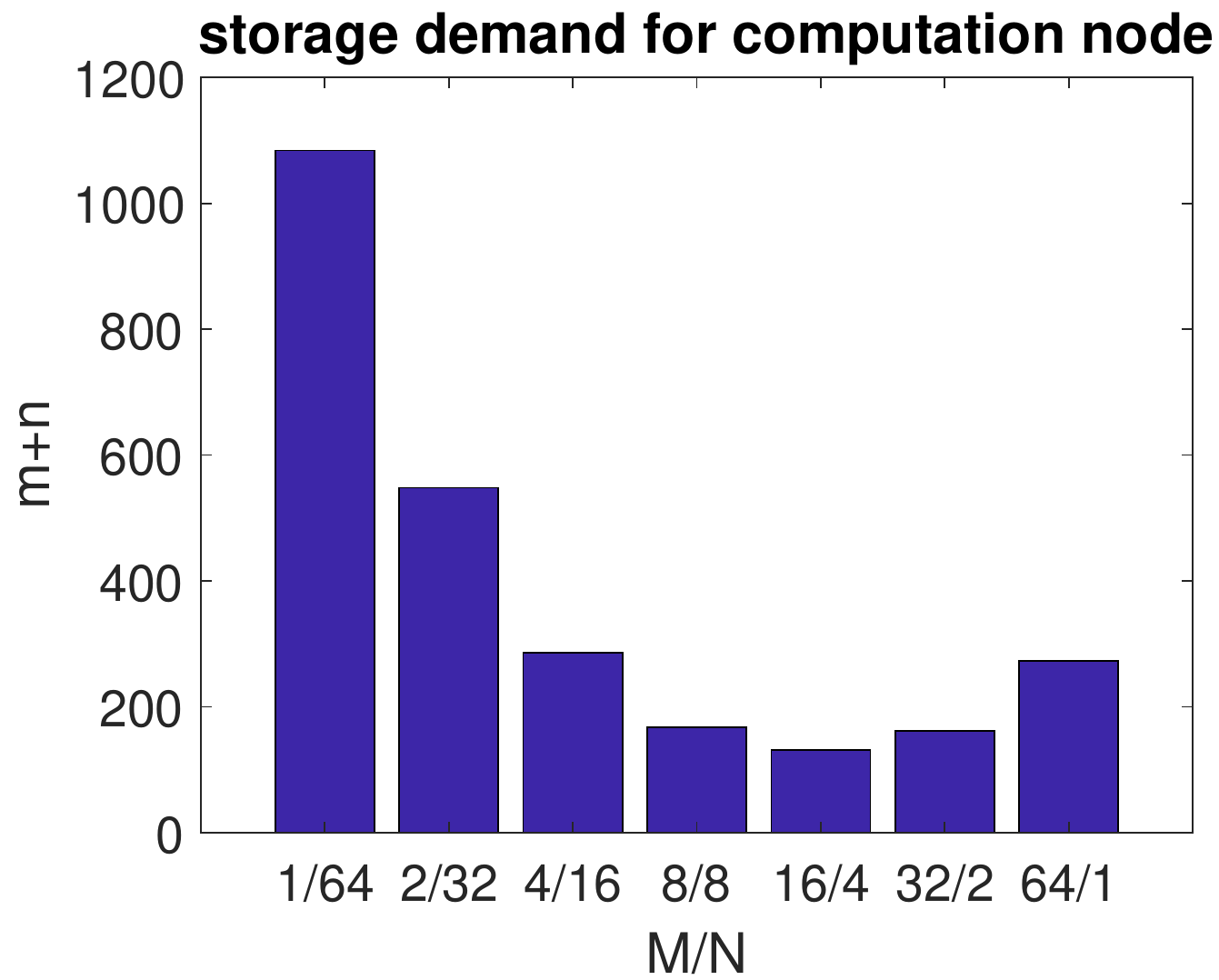}%
}

\subfloat[]{%
  \includegraphics[clip,width=70mm]{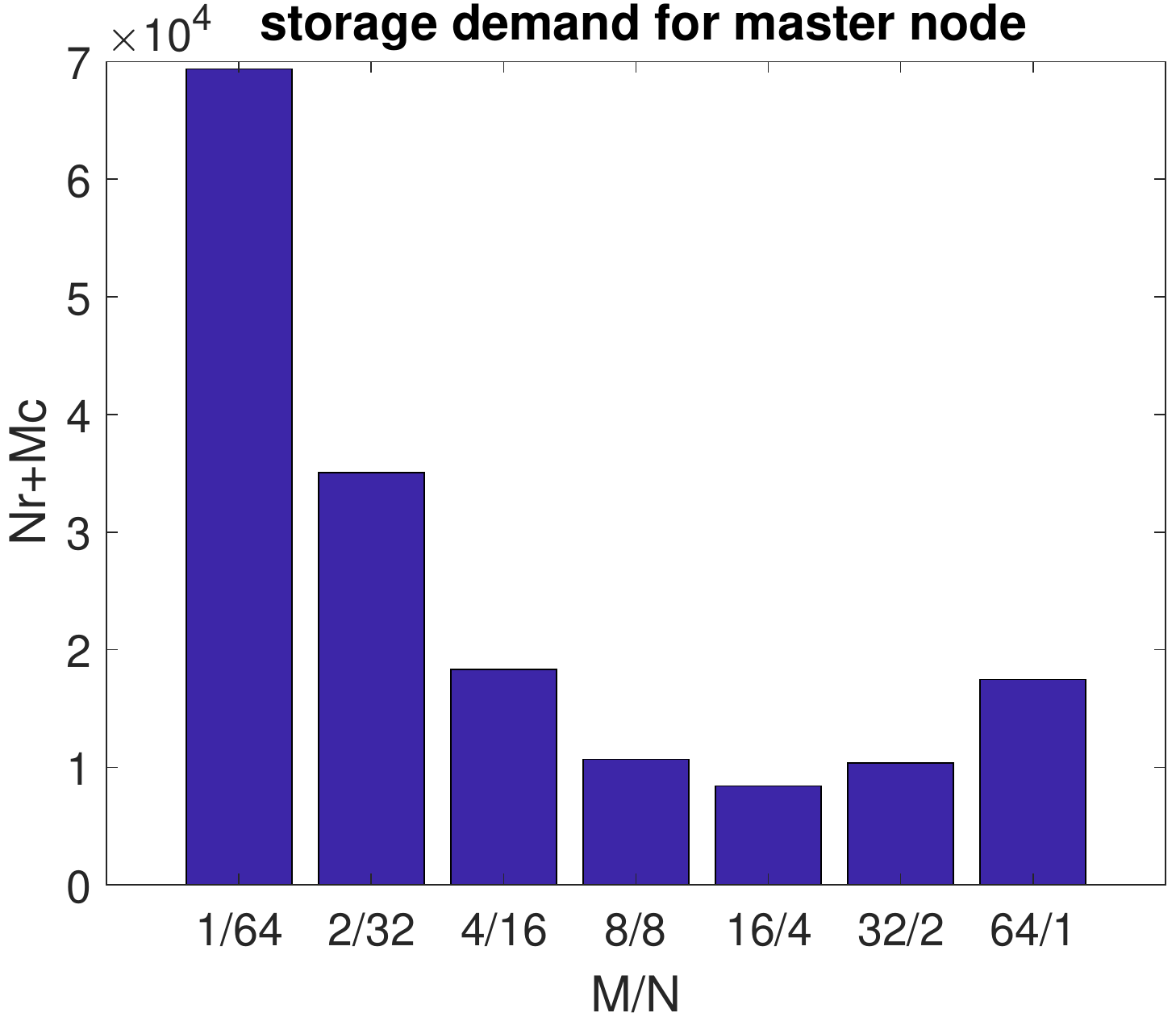}%
}
\caption{The computational complexity of the computations in each node is proportional to $mn$, where $m$ and $n$ are the numbers of rows and columns in$\A_I^J$. Storage requirements in each compute node is however proportional to $m+n$, which is shown here in panel (a). For the master node, storage requirements are proportional to $Mr+Nc$, which is shown in panel (b).}
\label{fig4}
\end{figure}

According to Eq.\ref{Eq1}, we divide $\A$ into different numbers of row and column blocks and compare convergence speed. The results  shown in Fig.\ref{CompaDiffMN} indicate that BSGD does not encourage the partition in the column direction.
Whilst this indicates that we do not want to partition in the column direction, different partitions lead to different storage demand. This is shown in Fig.\ref{fig4}, where we show the amount of storage that is required in the compute nodes and the master node.

\begin{figure}[htp]
\center
\subfloat[]{%
  \includegraphics[clip,width=70mm]{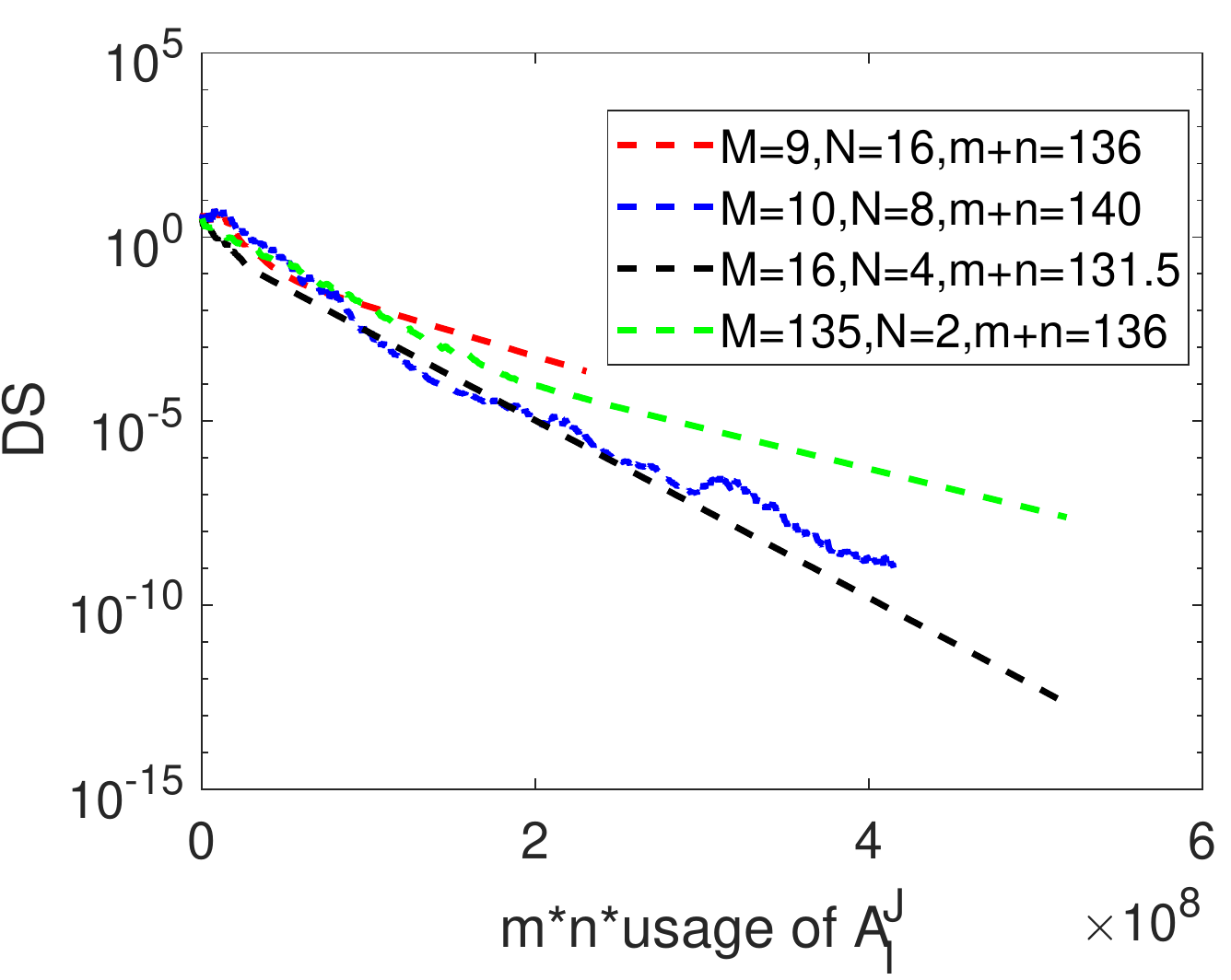}%
}

\subfloat[]{%
  \includegraphics[clip,width=70mm]{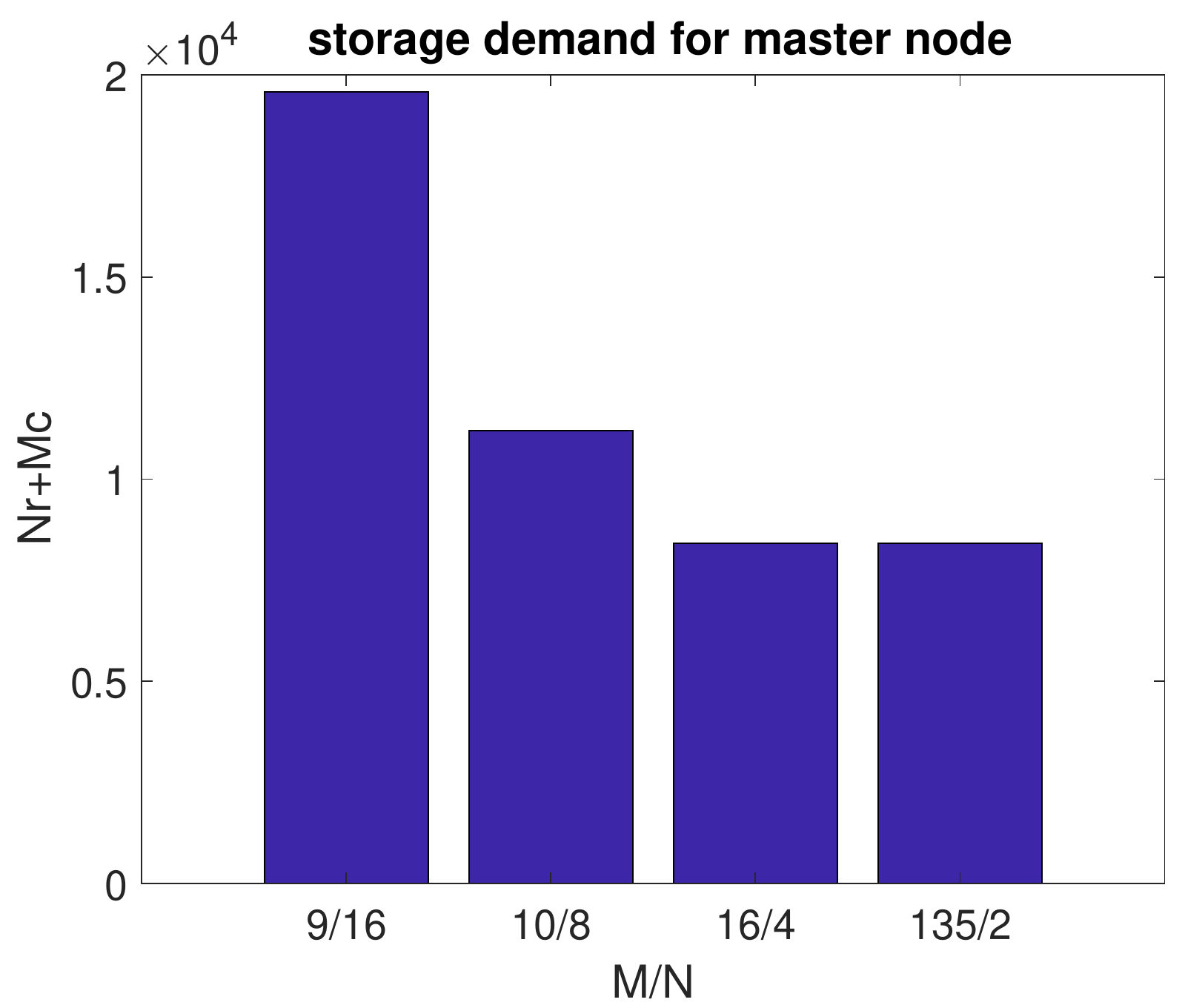}%
}
\caption{Convergence of BSGD over 3000 iterations (a) when we have 2 compute nodes. The best compromise between convergence speed and master node storage demand (b) is found for $M=16,N=4$.}
\label{CompareFixSto}
\end{figure}

Note that in the previous simulations, we kept the product $MN$ fixed as in this case, computation speed per iteration remains constant. However, when storage demand is the limiting factor, then the $m+n$ become limited ($m,n$ are rows and columns of $\A_I^J$). To explore this, we fix $m+n\leq 140$. The results, shown in  Fig.\ref{CompareFixSto} for different values of $M$ and $N$ are now plotted against the number of times we compute multiplications by matrices $\A_I^J$, multiplied by $mn$ to normalise computation time differences.
We here assume that there are only two nodes in the parallel network and the selection criteria for $\alpha$ and $\gamma$ follows Eq.\ref{Eq1}. The convergence speed and storage demand in the master node is shown in Fig.\ref{CompareFixSto}. Note that when $M=135,N=2$ (green line), BSGD becomes SAG. For large $N$, storage demand in the master node increases significantly. Luckily, convergence is the fastest for moderate values of $N$.

\subsection{Automatic parameter tuning}
\label{AutParaTun}
\begin{figure}[htb]
\centering     
\includegraphics[width=70mm]{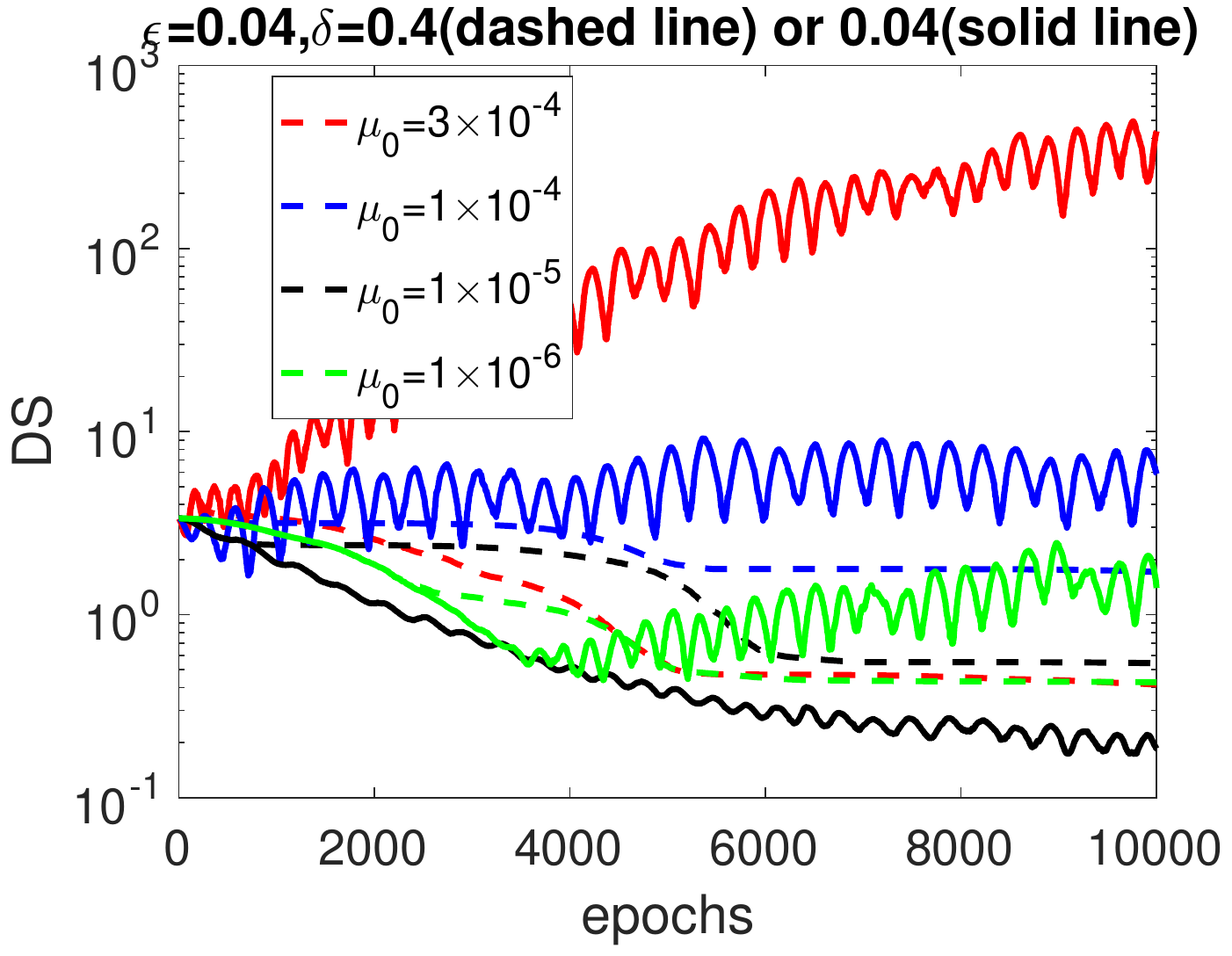}
\caption{$\mu_0$ is the initial step length. Different color stands for different $\mu_0$. Only using criteria 1 does not overcome issues with parameter selection. The dashed lines use a large $\delta$ leading to the algorithm getting stuck, whilst solid lines use a small $\delta$ leading to oscillation and divergence.}
\label{CompareFixSto2}
\end{figure}

We first demonstrate that criteria 1 ( line 8 in Algo \ref{cre1}) on its own is not sufficient for parameter tuning. If  $\delta$ is small, $\mu$ is not effectively reduced, whilst for large $\delta$, the $\mu$ tends to become too small and the iterations get `stuck'. Simulation results to prove this are shown in Fig.\ref{CompareFixSto2}.

\begin{figure}[htp]
\center
\subfloat[]{%
  \includegraphics[clip,width=70mm]{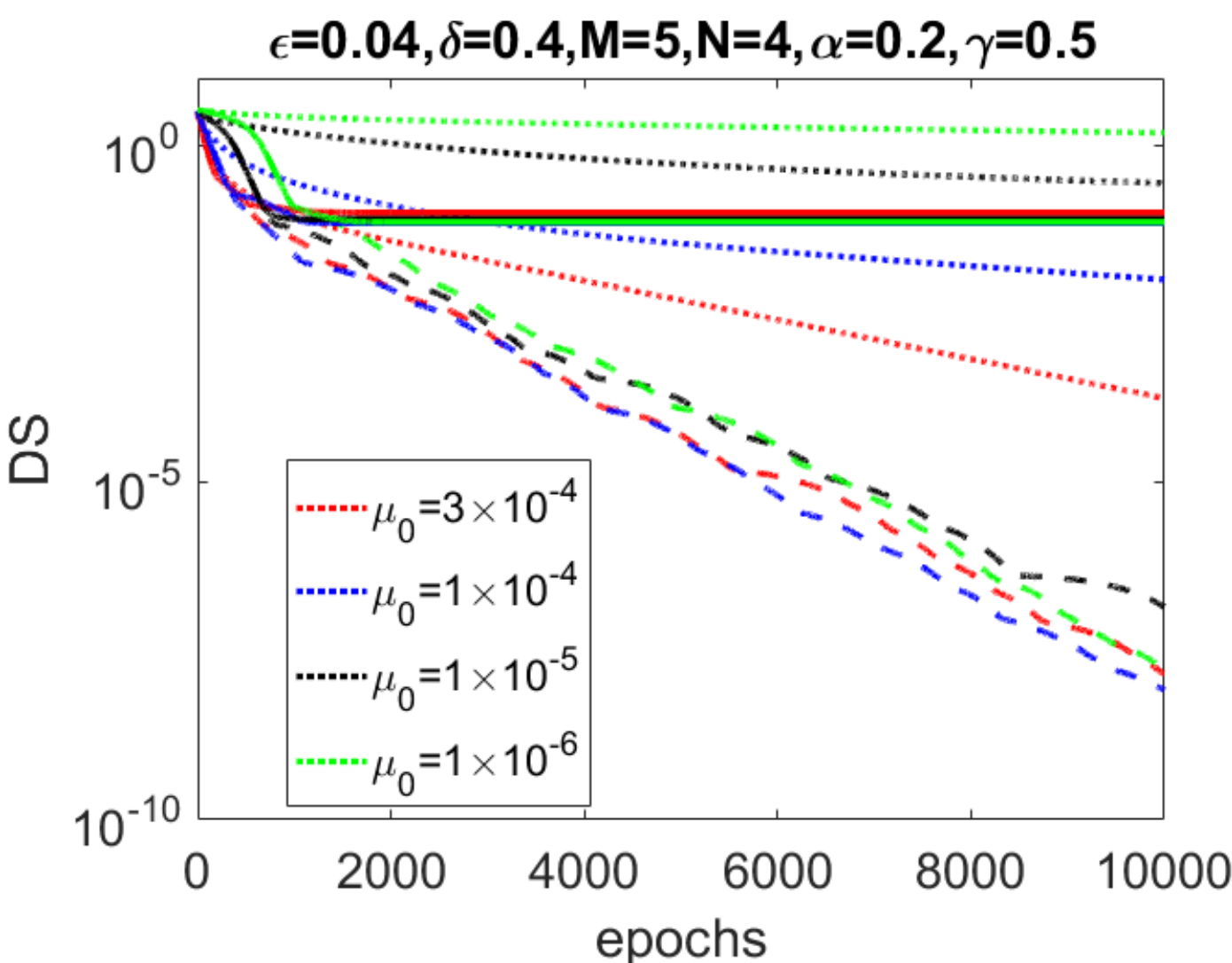}%
}

\subfloat[]{%
  \includegraphics[clip,width=70mm]{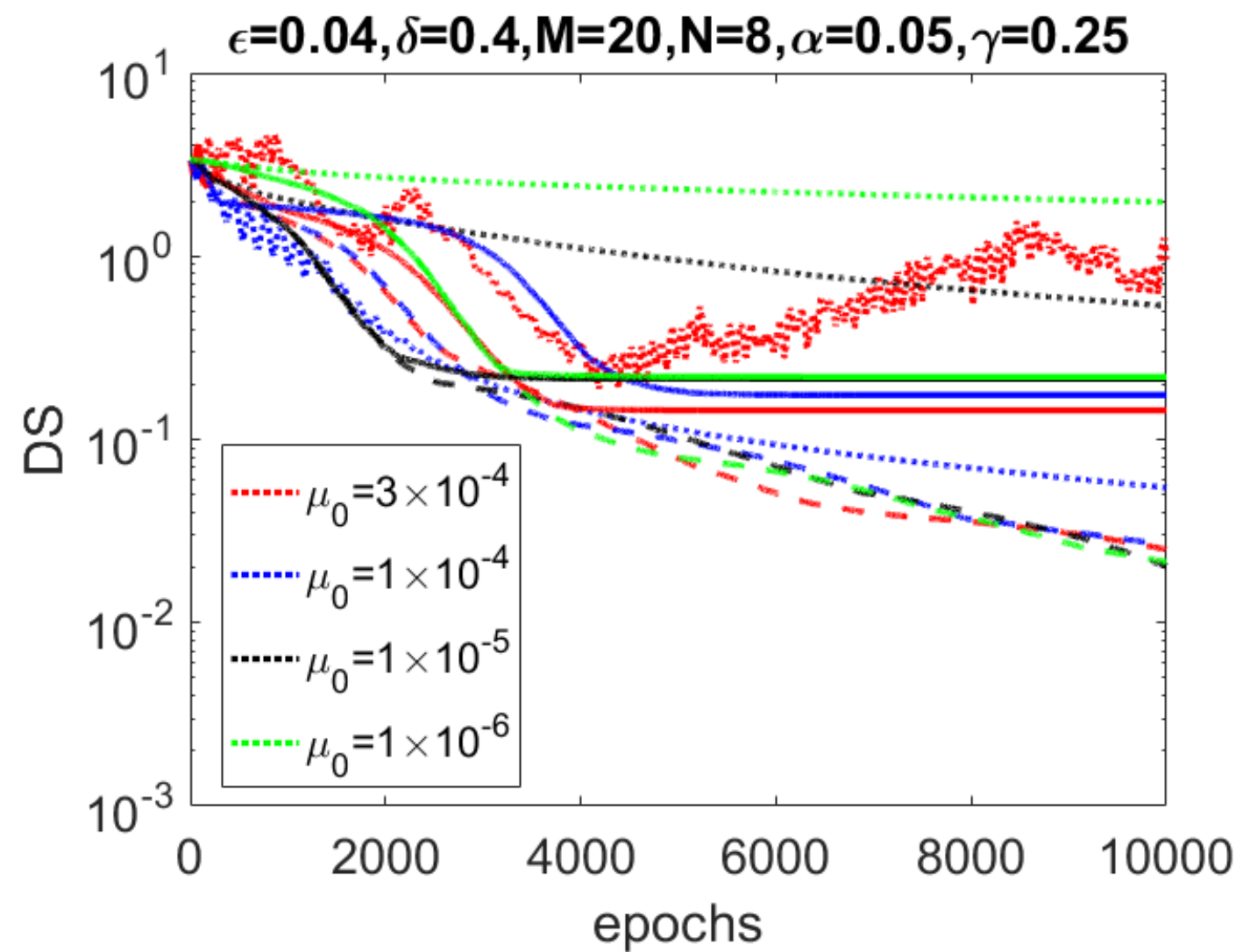}%
}
\caption{The dotted line is the original BSGD method with constant step length $\mu_0$. The solid line is the automatic parameter tuning method using criteria 1 and the dashed line is the automatic parameter tuning using criteria 1 and 2. Panels (a) and (b) show different scenarios in terms of $M$, $N$, $\alpha$ and $\gamma$.}
\label{auto}
\end{figure}

Automatic parameter tuning works if we combine criteria 1 with criteria 2 (line 9 in Algo \ref{cre1} ). We here use $\delta =0.4$. The compute nodes is still set as 2. The results, shown in Fig.\ref{auto}, demonstrate that automatic tuning (dashed lines) allows faster convergence than the original method (dashed line) and the method using criteria 1 only (solid line).

\subsection{Incorporating the TV constraint}
\label{D}
To explore Total Variation regularisation,  in Fig.\ref{Scanning geometry}, the $K$ is increased to 64. $OP$ and $OD$ is 100 and the detector has 180 elements. The scanning angle increment is $1^{\circ}$ and the point source only rotates through $180^{\circ}$.  In this section, the $\textbf{
Signal Noise Ratio}$ (SNR) of $\x$ is defined as $20\log_{10}\frac{\|\x_{true}\|}{\|\x_{dif}\|}$, where $\Vert\x_{dif}\Vert$ is the $\ell_2$ norm of the difference between reconstructed image vector and the original vector $\x_{true}$.  The $\lambda=0.1$ and the projection $\y$ are influenced by Gaussian noise, with an SNR (similar definition holds) of 28.8dB.

\begin{figure}[htb]
\centering     
\includegraphics[width=70mm]{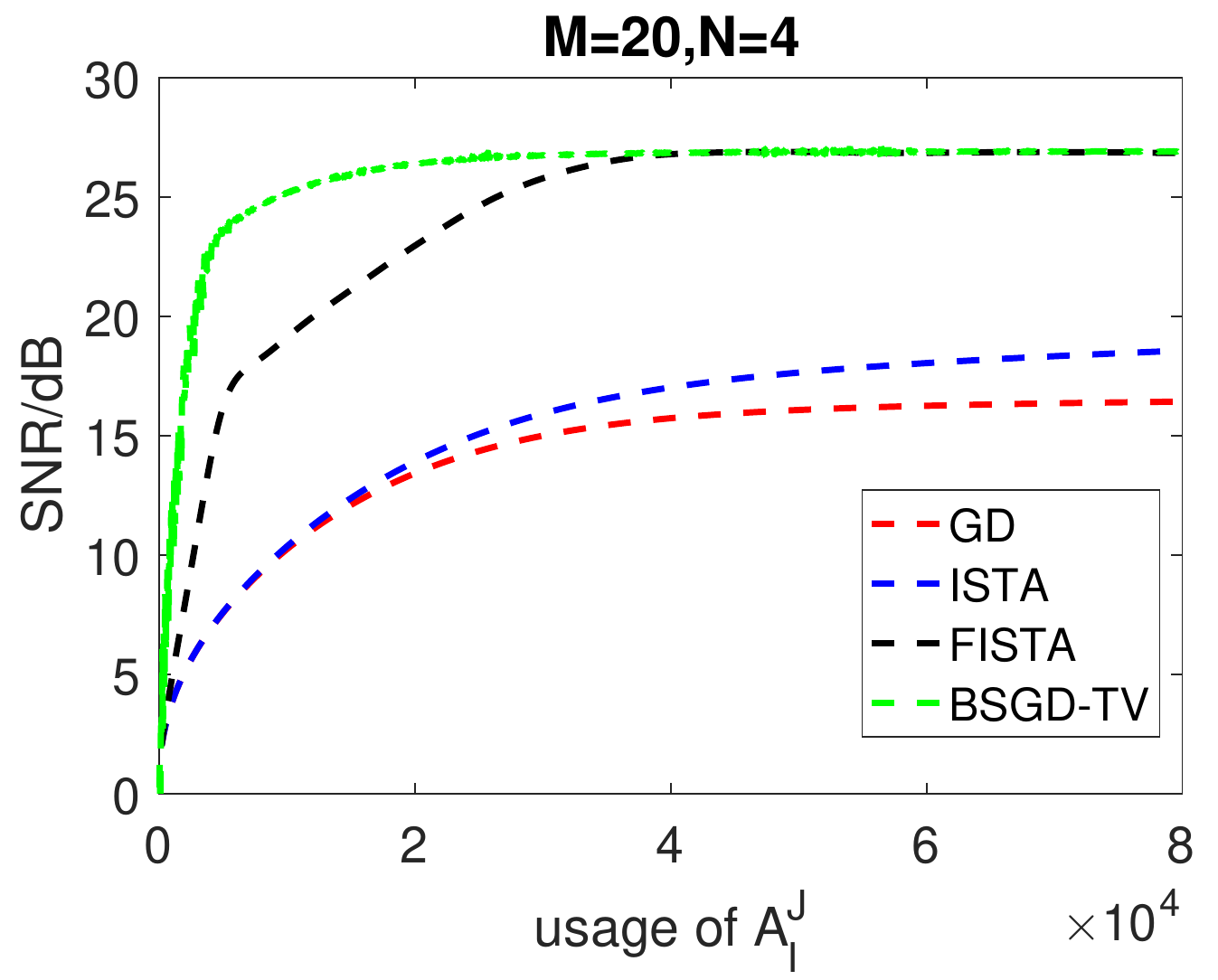}
\caption{Comparison between BSGD-TV, FISTA, ISTA and Gradient Descend (GD) (without TV constraint). The step length $\mu$ for ISTA, GD and BSGD-TV is 0.0004. BSGD-TV uses $M=20,N=4,\alpha=0.05 and \gamma=0.5$ with 2 nodes available in the network.  $\lambda$ in Eq.\ref{TVequ} is $0.1$. The low SNR of GD suggests the necessity of incorporating the TV norm here. It can be seen that the BSGD-TV converge faster than FISTA methods.}
\label{Compare2}
\end{figure}

\begin{figure}[htp]
\center
\subfloat[GD]{%
  \includegraphics[clip,width=42mm]{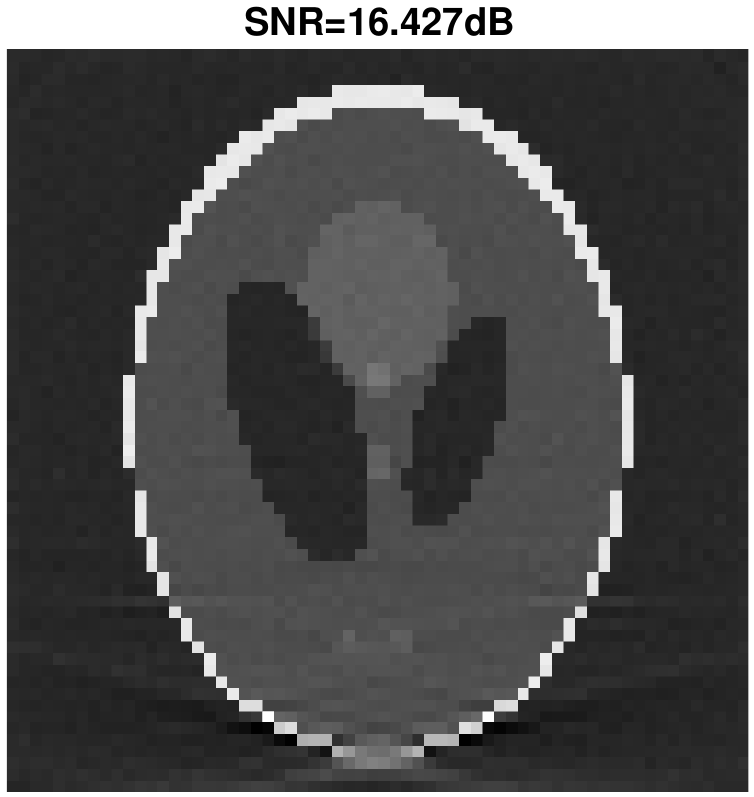}%
}
\subfloat[ISTA]{%
  \includegraphics[clip,width=42mm]{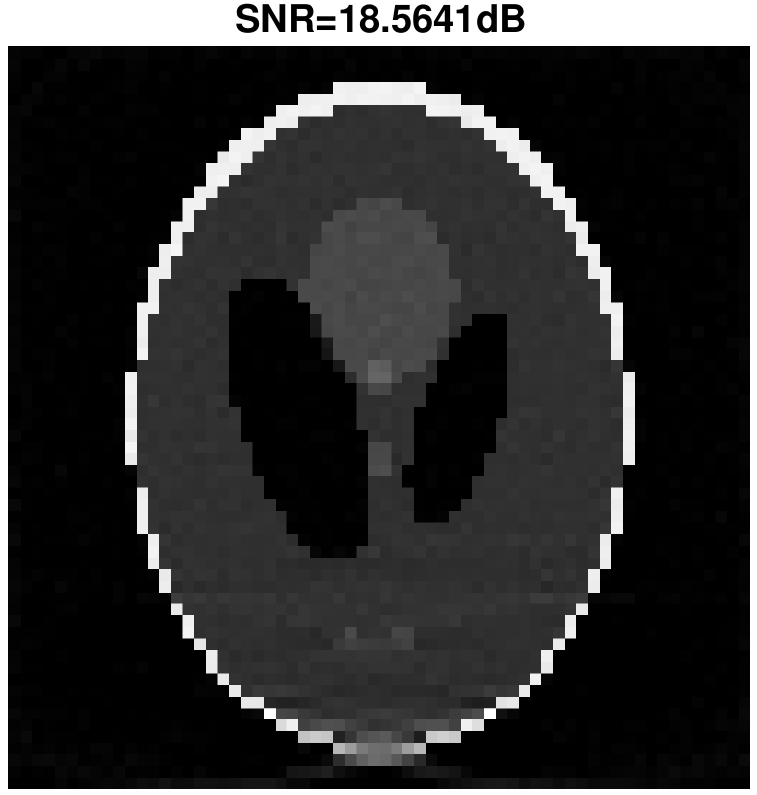}%
}

\subfloat[FISTA]{%
  \includegraphics[clip,width=42mm]{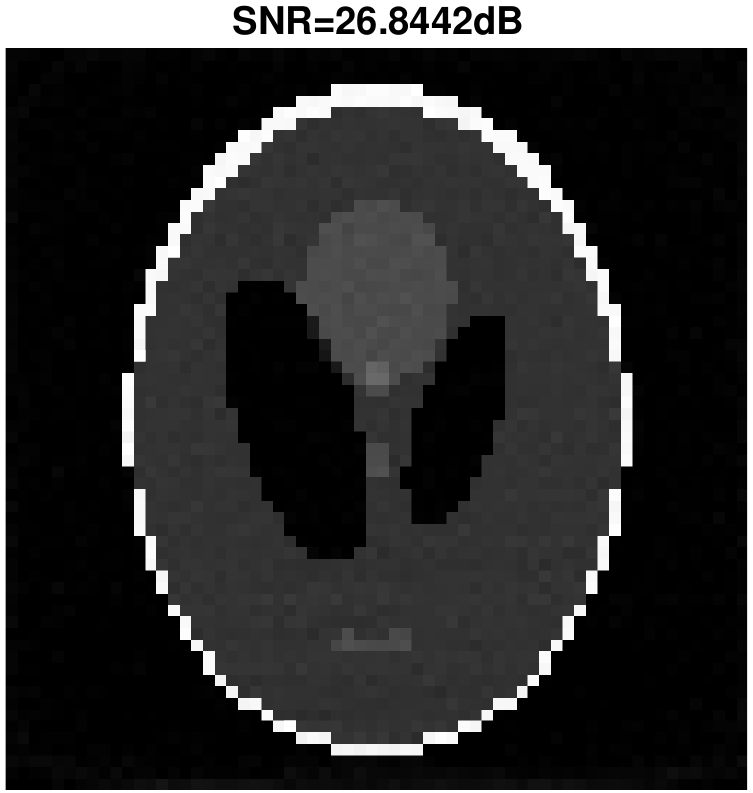}%
}
\subfloat[BSGD]{%
  \includegraphics[clip,width=42mm]{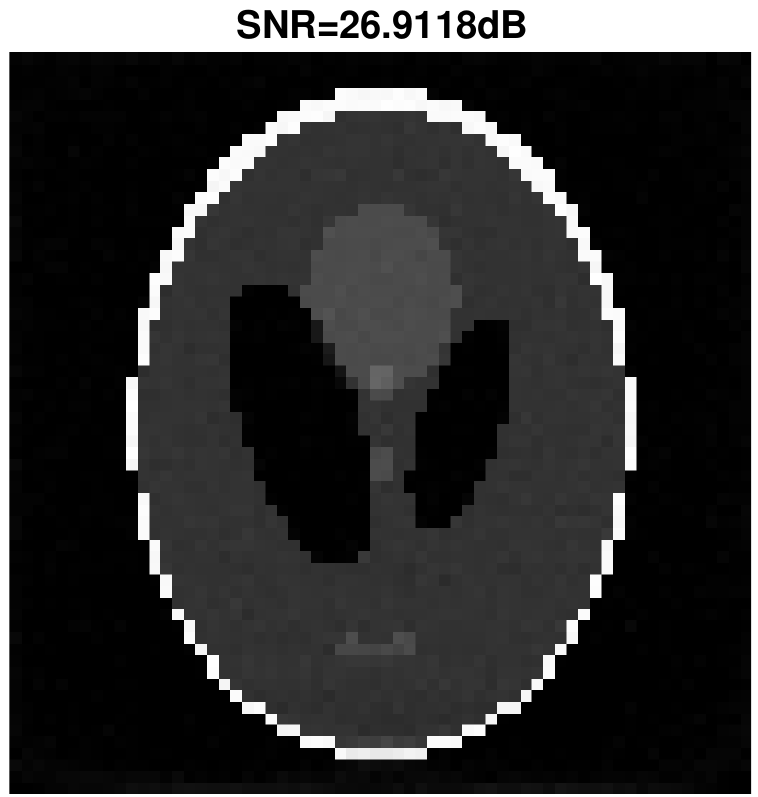}%
}
\caption{Reconstruction results after 500 effective epochs.}
\label{figcompare2}
\end{figure}

The change of SNR is shown in Fig.\ref{Compare2}. 
We here plot SNR against effective epochs, where an effective epoch is a normalized iteration count that corrects for the fact that the stochastic version of our algorithm only updates a subset of elements at each iteration.Comparing BSGD-TV against FISTA and ISTA, BSGD-TV  shows a faster convergence speed, even though it only update blocks of $\x$ in each iteration and only applies the TV-based de-noising after a period of iterations. A visual comparison after a fixed number of effective epochs is shown in Fig. \ref{figcompare2}.

Note that ADMM-TV \cite{boyd2011distributed} is another algorithm that can be distributed over several nodes, however, our previous work has demonstrated that ADMM-TV is significantly slower than  TV constraint version of CSGD \cite{gao2018joint} and thus has not been included in the comparison here. 

\subsection{Applying BSGD in 3D CT reconstruction}
\label{E}
In this section, a workstation containing two NVIDIA GEFORCE GTX 1080Ti GPUs is adopted to demonstrate BSGD's performance on realistic data sizes. We used a 3D cone beam scanning geometry similar to the 2D simulation as defined in Fig.\ref{Scanning geometry}. In simulation, $OP=1536$ mm, $OD=1000$ mm and the detector is a square panel with side length($EF$) of 400 mm. The reconstruction volume is a cube with side length of 256 mm. The point source and the centre of the square detector are located at the middle slice of the 3D volume. They rotate around the volume horizontally for a full circle with angular increments of $1^\circ$. We test BSGD for increasing data sizes (see Table \ref{Table_Parameter}) and compare it to other  methods.
\begin{table}[htbp]
  \centering
  \caption{Three different reconstruction scales}
    \begin{tabular}{cccc}
    \toprule
          & case1 & case2 & case3 \\
    \midrule
    Detector & 400*400 & 1000*1000 & 2000*2000 \\
    \midrule
    Volume &  256*256*256  & 512*512*512 & 1024*1024*1024 \\
    \midrule
    Rows of $\mathbf{A}$  & $5.76\times 10^7$ &$3.6\times 10^8$ & $1.44\times 10^9$ \\
   \midrule
  Columns of $\mathbf{A}$  &$1.68\times 10^7$ & $1.34\times 10^8$ & $1.07\times 10^9$ \\
    \bottomrule
    \end{tabular}%
  \label{Table_Parameter}%
\end{table}%
 The simulations are performed in MATLAB R2016b together with a purpose built version of the TIGRE toolkit \cite{biguri2016tigre} that uses OpenMP to synchronize the two GPUs, performing two FPs or two BPs simultaneously. Simulation results show that BSGD with importance sampling and automatic parameter tuning can be applied in realistic settings and is faster than existing methods in a multi-GPU work station. 

\subsubsection{Time required for FP, BP and other operations}
We start by looking at the time required to compute FB/BP and contrast this time to the other computational overheads of the method. 
For one GPU, we process two data blocks one after the other whilst for two GPUs two blocks are computed in parallel. We here divided the image into 8 cubic blocks ($N=8$) and randomly partitioned the 360 projections into 5 groups ($M=5$). As we here look at computation speed of individual FP and BPs, no noise was added to the projections. 

\begin{figure}[htp]
\center
\subfloat[FP]{%
  \includegraphics[clip,width=70mm]{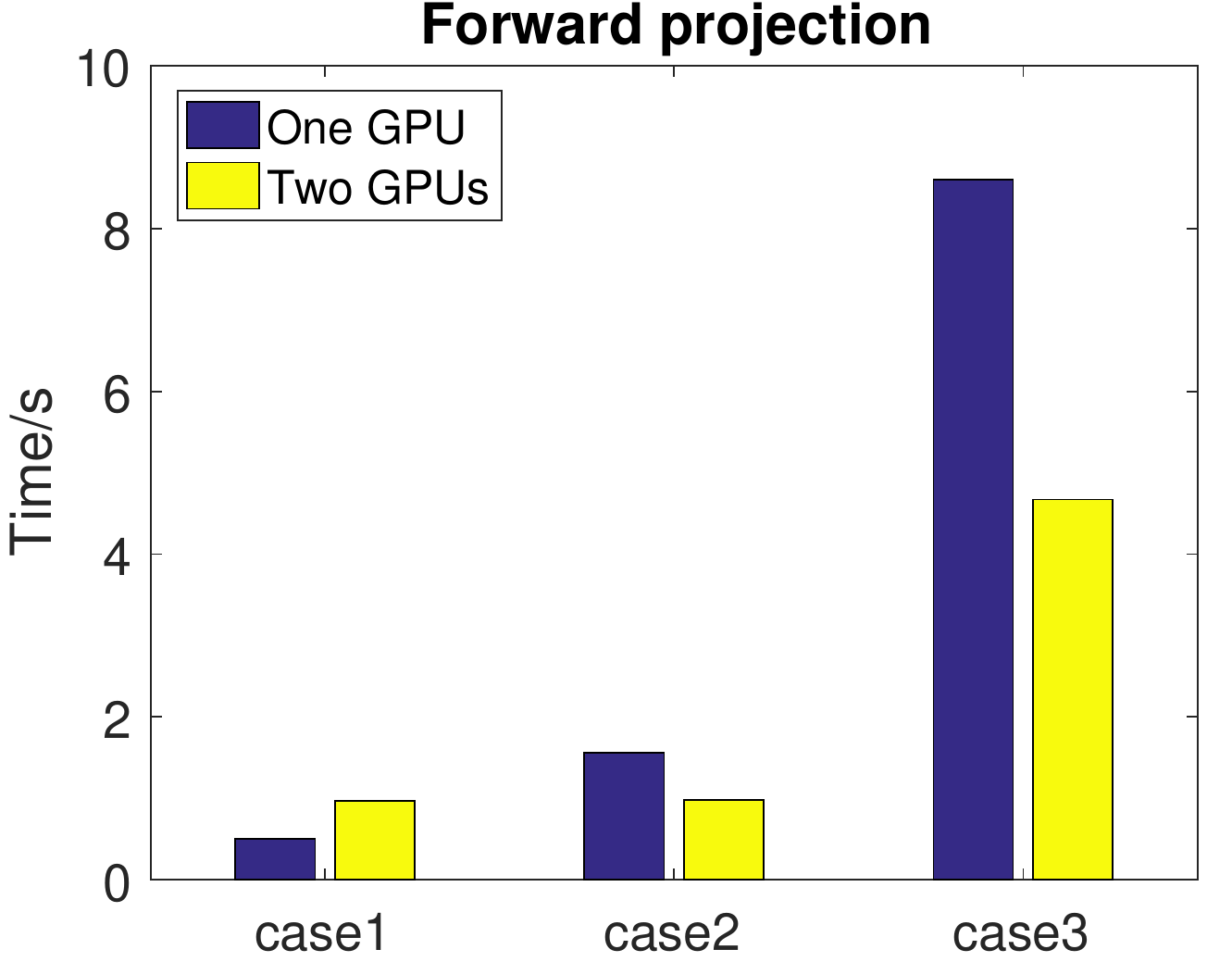}%
}

\subfloat[BP]{%
  \includegraphics[clip,width=70mm]{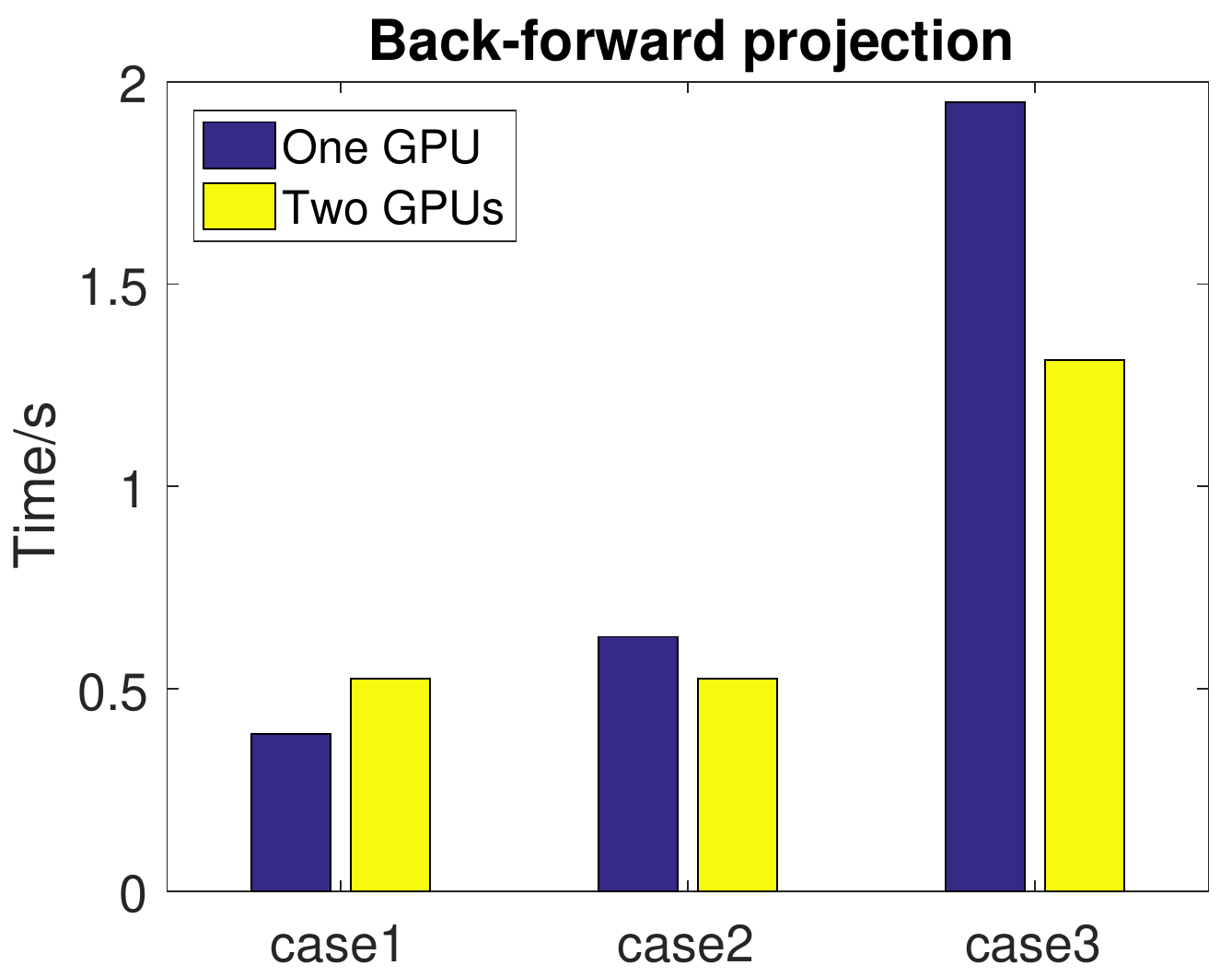}%
}
\caption{When the problem size is small, using two GPUs does not provide acceleration as the overhead on communication and synchronization dominates. However, for realistic scales, using two GPUs nearly achieves the optimal doubling of computation speed with two GPUs for the FP and 2/3 acceleration of the BP.}
\label{ProjectionCompare}
\end{figure}

The overall time spent on FP and BP per iteration are shown in Fig.\ref{ProjectionCompare}. The measurements here are averaged over 10 repeated runs. 

\begin{figure}[htp]
\center
\subfloat[Using one GPU]{%
  \includegraphics[clip,width=70mm]{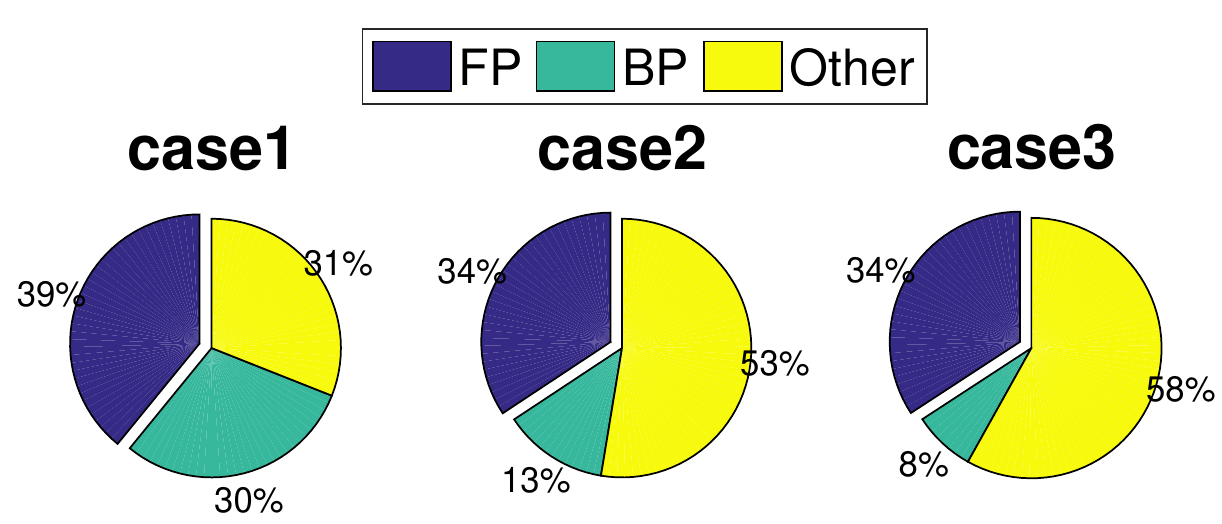}%
}

\subfloat[Using two GPUs]{%
  \includegraphics[clip,width=70mm]{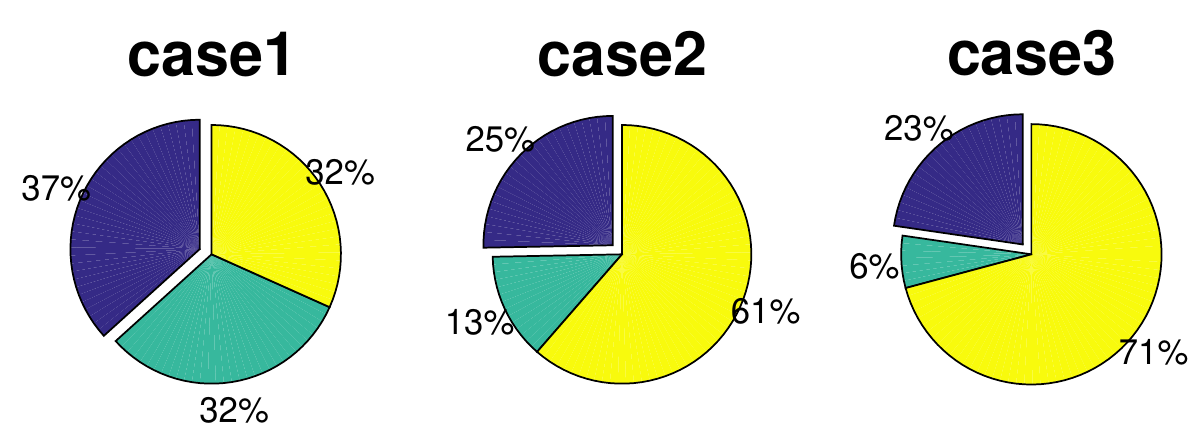}%
}
\caption{The relative time spent on each stage during an entire reconstruction task. For increased problem sizes, the percentage of time spent on FP and BP decreases faster when using two GPUs.}
\label{percentage}
\end{figure}

We also measured the proportion of time each iteration of BSGD spent on FP and BP during reconstruction (See Fig.\ref{percentage}), which shows that for increasing problem sizes, FP and BP become increasingly smaller fractions of overall cost. As data transfer and other operations are similar in the 1 and 2 GPU settings, this further demonstrates that multi-GPU reconstruction is beneficial to reduce the time spent on FP/BP. 

\subsubsection{Quality of reconstruction}
\begin{figure}[htp]
\center
\subfloat[SNR trend]{%
  \includegraphics[clip,width=70mm]{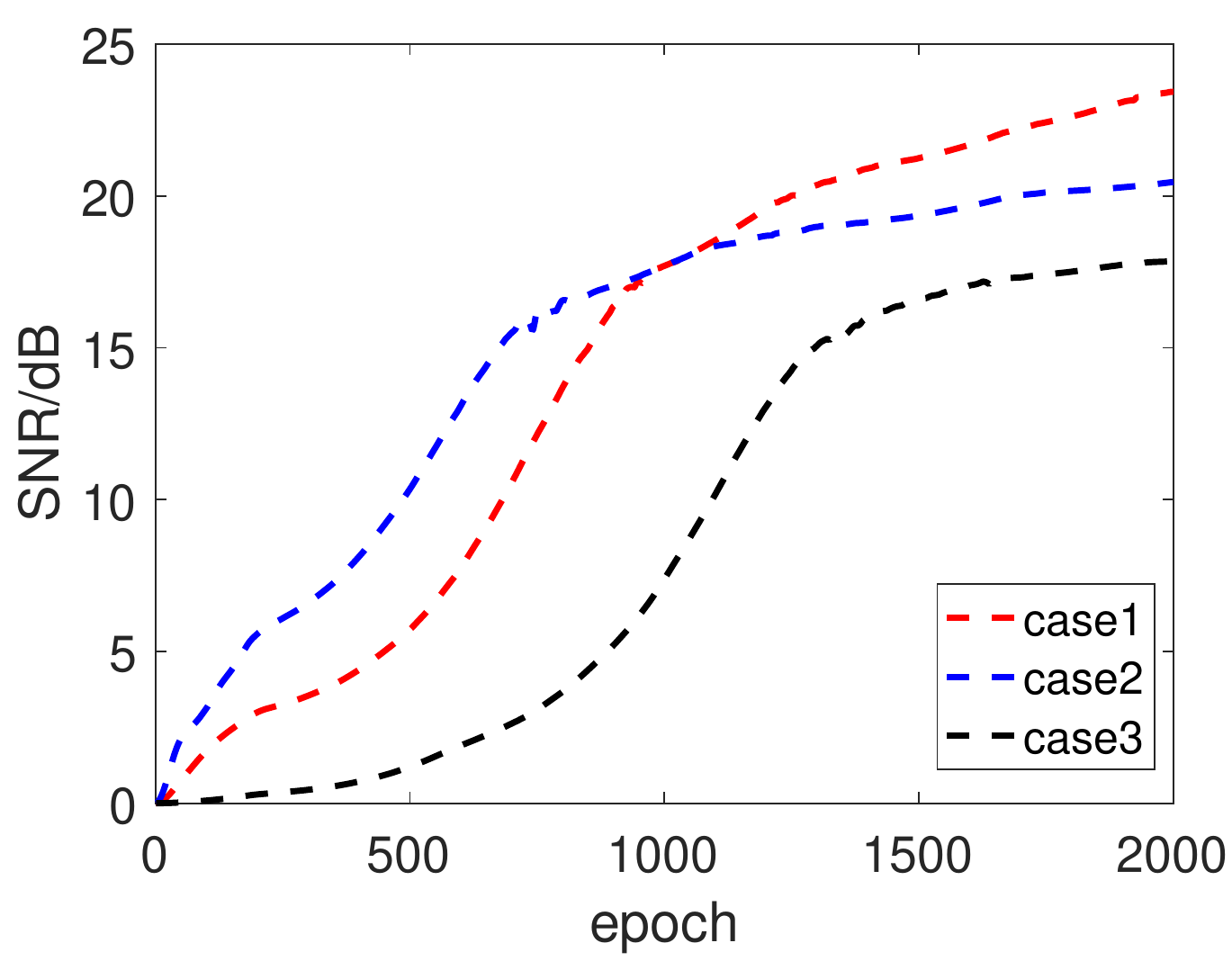}%
}

\subfloat[step length trend]{%
  \includegraphics[clip,width=70mm]{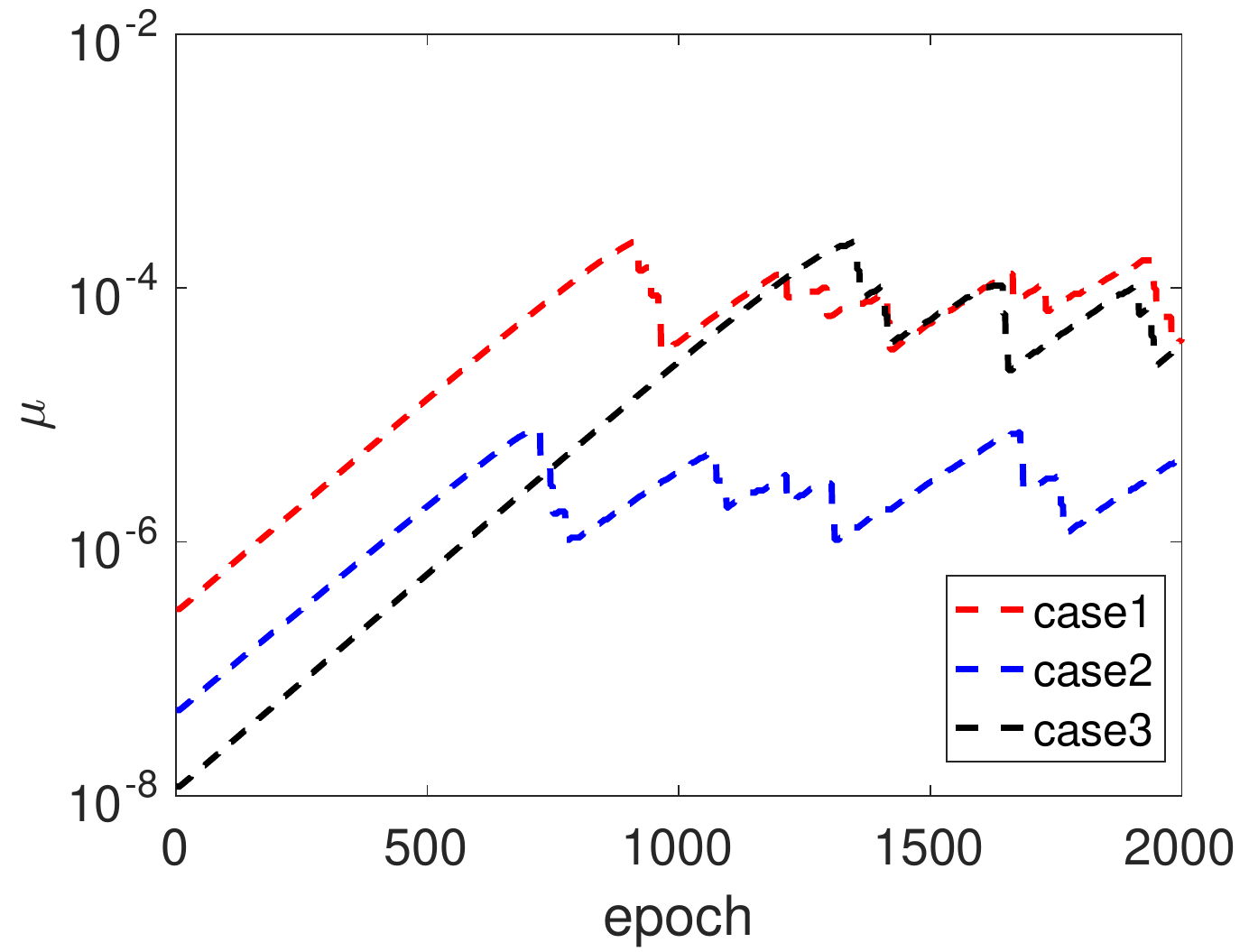}%
}
\caption{During 2000 epochs, BSGD provides an increasing SNR trend under different data-set scales. The step length can also be automatically adjusted into a range that enable the iteration results move towards to the true solution.}
\label{FigSnrLarge}
\end{figure}

We next look at the quality of reconstruction for the three scenarios. Since there are two GPUs are available, we thus set $M=5,N=8,\alpha=\frac{1}{5}$ and $\gamma=\frac{2}{8}$. We here measure quality in terms of SNR. The results are shown in Fig.\ref{FigSnrLarge}. 

\begin{figure}[htb] 
  \centering 
    \includegraphics[width=80mm]{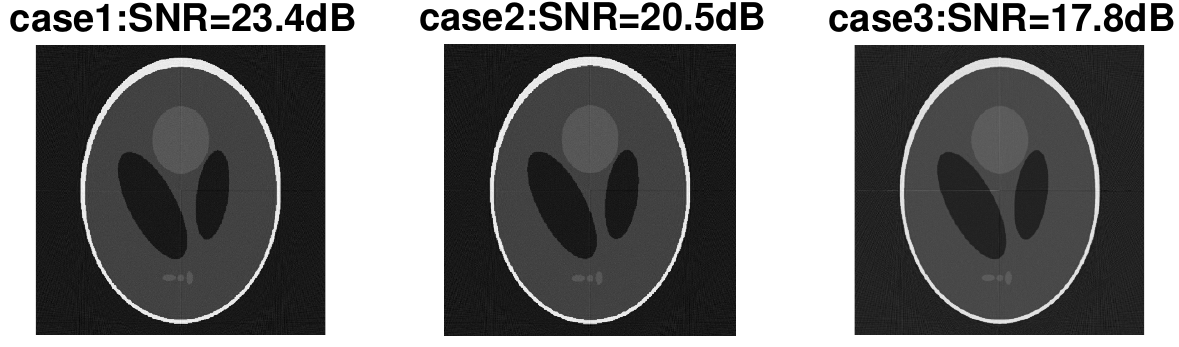} 
  \caption{Slices from the 3D reconstruction for different problem dimensions, showing the effectiveness of BSGD under different cases.}  
  \label{figsubfigIma} 
\end{figure}

Reconstructed slices are shown in Fig.\ref{figsubfigIma}. 

\subsubsection{Comparison with other methods}
In this section, we demonstrate the advantage of BSGD compared with other methods. A cubic skull skeleton provided by the TIGRE toolkit was used and reconstructed using different methods. The object to detector distance was 536 mm, source to object distance was 1000 mm and we again collected 360 equally spaced projections and reconstructed onto a 256 by 256 by 256 grid. The detector used $512 \times 512$ pixels. The side length of each voxel and detector pixel were 1 mm, so that  $\A\in\R^{9.4\times 10^7* 1.7\times 10^7}$. The projection data $\y$ is deteriorated by white Gaussian noise with SNR of 28.1 dB.  In our simulations, we assume that each computation node (GPU) can only process $\frac{1}{8}$ of the volume and 18 projections. Since in the previous simulation we have demonstrated that BSGD can outperform CAV and SIRT, we here compared BSGD-IM with SAG, SVRG, GD, GD-BB\cite{barzilai1988two}, FISTA and ORBCDVD \cite{wang2014randomized}. Except for BSGD-IM, the other methods divide $\A$ into $20*8$ sub-matrices (i.e. $M=20,N=8$ to enable each row blocks contain 18 projections and each column blocks contain ${\frac{1}{8}}^{th} $volume. $\alpha=\frac{1}{20},\gamma=\frac{1}{4}$ is set according to Eq.\ref{Eq1}) and consecutively process the sub-matrices on each GPU one at a time. BSGD-IM, sampling $\frac{1}{4}^{th}$ projection from each projection angle,  allows us to divide $\A$ into $5*8$ sub-matrices with $\alpha$ and $\gamma$ set to 0.2 and 0.5. By this division, it is guaranteed that the computation amount for FP and BP of BSGD-IM are the same with the other methods. This is because that despite that the BSGD-IM process 72 projection angles each time, for each projection angle, only $\frac{1}{4}^{th}$ projection data are used for each projection angle. As a result, the actual projection data size is equivalent to $72*\frac{1}{4}=18$ full projection angles, which is the size for other methods.

\begin{figure}[htb]
\centering     
\includegraphics[width=70mm]{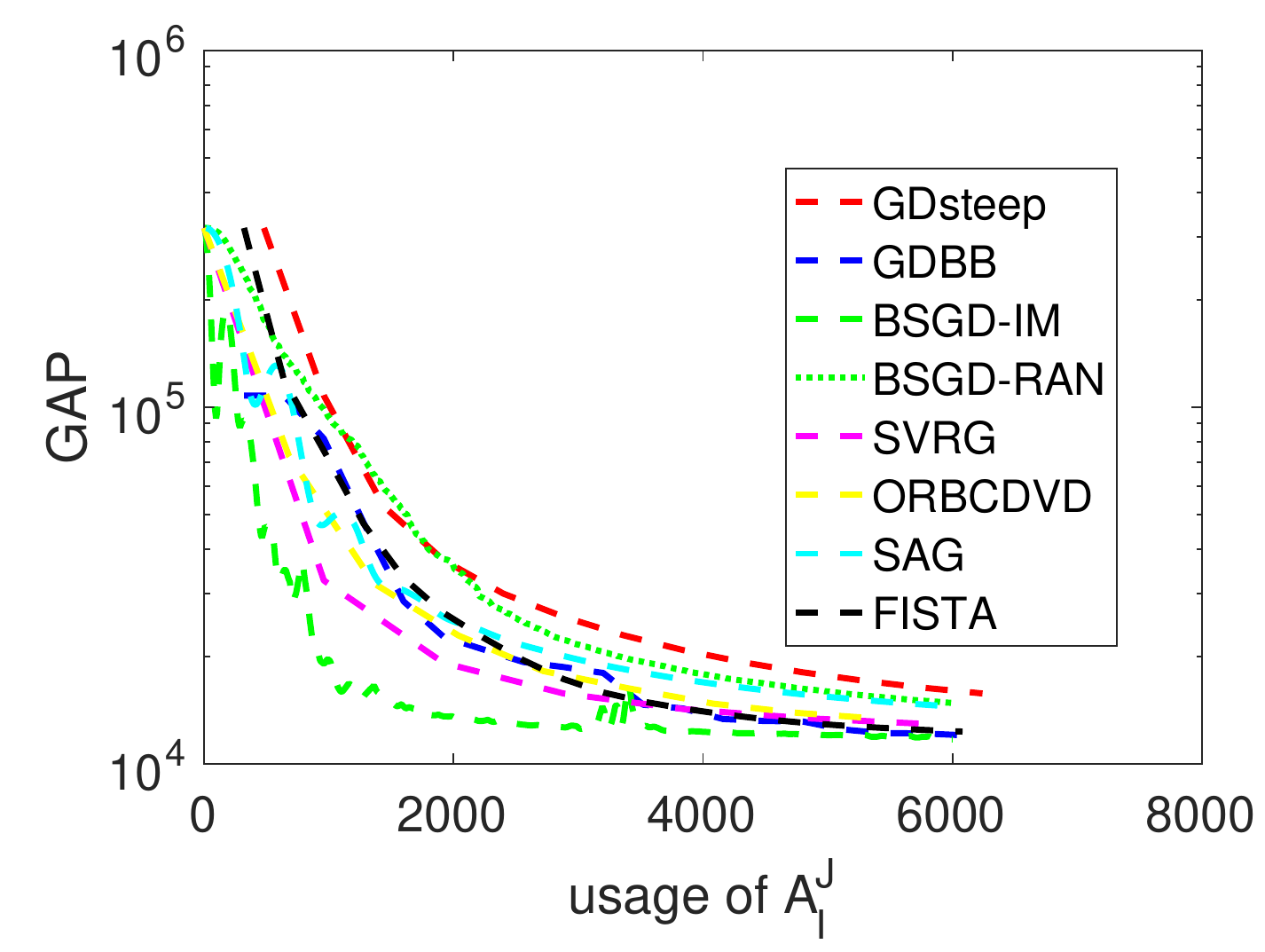}
\caption{BSGD-RAN is similar to BSGD-IM, but it uniformly selects the sub-projections at each projection angle while BSGD-IM selects sub-projections based on sub-matrix sparsity. Both BSGD methods use automatic parameter tuning. The parameters in the other methods were optimised to ensure optimal performance.}
\label{Compare}
\end{figure}

\begin{figure}[htp]
\center
\subfloat[Original slice]{%
  \includegraphics[clip,width=30mm]{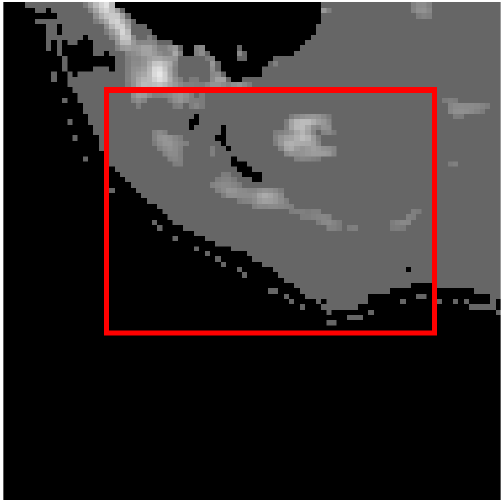}%
}
\subfloat[BSGD]{%
  \includegraphics[clip,width=30mm]{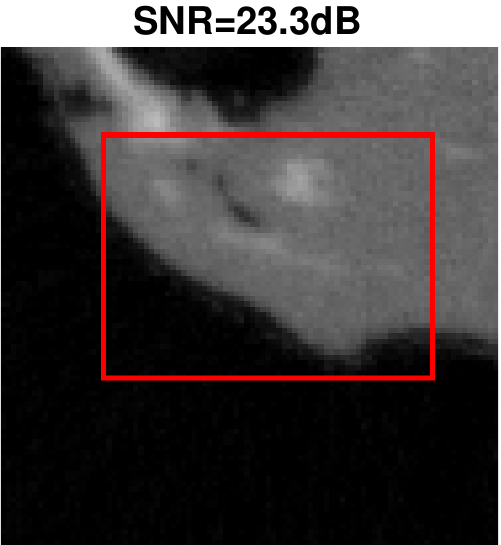}%
}
\subfloat[SVRG]{%
  \includegraphics[clip,width=30mm]{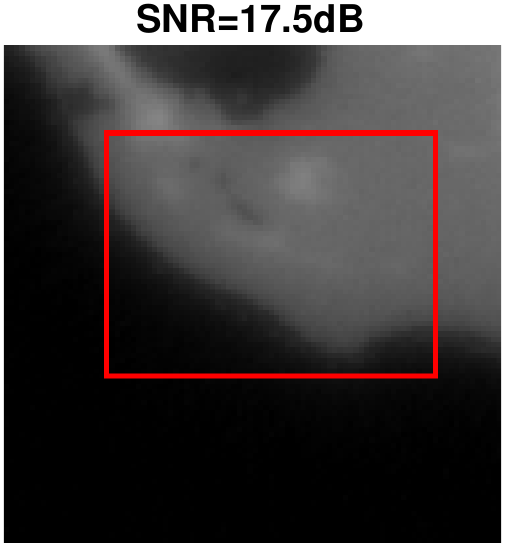}%
}

\subfloat[SAG]{%
  \includegraphics[clip,width=30mm]{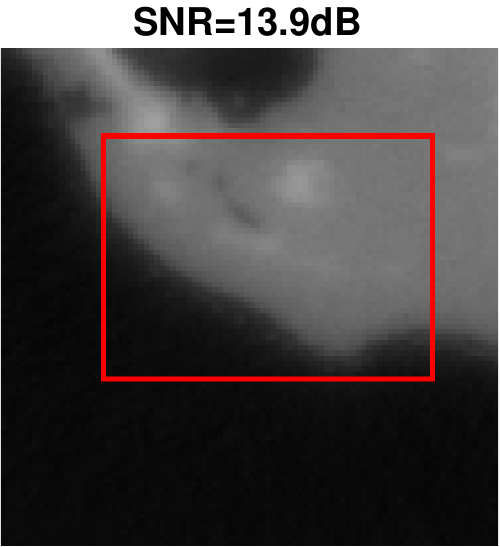}%
}
\subfloat[ORBCDVD]{%
  \includegraphics[clip,width=30mm]{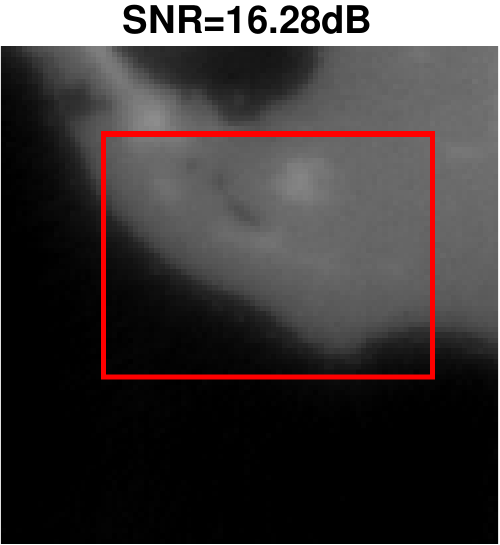}%
}

\caption{Reconstruction results after 2000 projections. BSGD provides better reconstructions with a higher signal noise ratio(SNR).}
\label{figcompare}
\end{figure}

In the large scale reconstruction case, calculating the least square solution itself can be time consuming, thus using the term $DS$ is inapplicable in realistic case. To reflect the speed of the iteration result approaching to the least square solution, we thus plot $GAP=\|\y-\A\x_{est}\|$  as a function of the number of usages of $A_I^J$ for each FP and BP. This is reasonable since the least square solution minimizes $GAP$ and a faster downward trend suggests a faster reconstruction speed.  Convergence results are shown in Fig.\ref{Compare}.
It can be seen that  BSGD-IM is faster than the other methods. Reconstruction results are shown in Fig.\ref{figcompare}, where we show a subsection of a 2D slice after 2000 forward and backward projections.

\section{Fixed point analysis}
In this part we show that BSGD has a single fixed point at the least square solution of the optimisation problem. The system matrix $\A\in\mathbb{R}^{r*c}$, has $M$ row blocks and $N$ column blocks. We vectorise the sets $\{\z^j\}_{j=1}^N$ and $\{\g^i\}_{i=1}^M$ and put their elements into the vector $\overline\z\in\mathbb{R}^{Nr*1}$ and $\overline\g\in\mathbb{R}^{Mc*1}$.
$\overline\A\in \mathbb{R}^{Nr*c}$ is a deformation of $\A$, defined as:
\begin{equation}
\overline\A = \begin{bmatrix}
\A_{I_1}^{J_1} &  \mathbf{0}&  \hdots&  \mathbf{0}\\
\vdots   &   \vdots   &         & \vdots \\
\A_{I_M}^{J_1} & \mathbf{0} & \hdots & \mathbf{0}\\
\mathbf{0} & \A_{I_1}^{J_2} & \hdots & \mathbf{0}\\
\vdots & \vdots & & \vdots \\
\mathbf{0} & \A_{I_M}^{J_2} & \hdots & \mathbf{0}\\
\mathbf{0} & \mathbf{0} & & \A_{I_1}^{J_N}\\
\vdots &\vdots & &\vdots\\
\mathbf{0} & \mathbf{0}&\hdots & \A_{I_M}^{J_N}
\end{bmatrix}
\end{equation}
$\overline{\A^T}\in\mathbb{R}^{Mc*r}$ is a similar deformation of $\A^T$.
Let us also introduce the matrix $\overline{\I}_{Nr}=[\I_r,\I_r,\hdots,\I_r]\in\mathbb{R}^{r*Nr}$, where we concatenate $N$ identity matrices $\I_r$ each of size $r*r$. With this notation we obtain:
\begin{equation}
\begin{aligned}
& \overline{\I}_{Nr}\overline{\A}=\A,\\
& \overline{\I}_{Mc}\overline{\A^T}=\A^T,\\
&  \sum_{j=1}^N\z^j=\overline{\I}_{Nr} \overline{\z},\\
& \g = \overline{\I}_{Mc} \overline{\g},\\
\end{aligned}
\label{overlineA}
\end{equation}
where the definition of $\g$ and $\{\z^j\}$ can be found in the algorithm description.
To encode the random updates over subsets of index pairs $I_i,J_j$, we introduce the random matrices $\bR_1\in \mathbb{R}^{Nr*Nr}$, $\bR_2\in \mathbb{R}^{Mc*Mc}$ and $\bR_3\in \mathbb{R}^{c*c}$ , which are diagonal matrices whose diagonal entries are either $0$ or $1$. With this notation, we can write the update of $\z$, $\g$ and $\x$ as:
\begin{equation}
\overline{\z}^{k+1} = \overline{\z}^k + \bR_1 \left[ \overline\A\x^k-\overline{\z}^k \right], 
\label{Updaz}
\end{equation}
\begin{equation}
\overline{\g}^{k+1} = \overline{\g}^k + \bR_2 \left[ \overline{\A^T} \left(\y-\overline{\I}_{Nr}\overline{\z}^{k+1}\right)-\overline{\g}^k \right]
\label{Updag}
\end{equation}
and 
\begin{equation}
\x^{k+1} = \x^k + \mu\bR_3 \overline{\I}_{Mc} \overline{\g}^{k+1},
\label{Updax}
\end{equation}
where $k$ is the epoch number. Inserting Eq.\ref{Updaz} into  Eq.\ref{Updag} and then Eq.\ref{Updag} into Eq.\ref{Updax}, the recursion in Eq.\ref{EquMat} is obtained:
\begin{equation}
\begin{bmatrix}
\overline{\z}^{k+1} \\
\overline{\g}^{k+1} \\
\x^{k+1} 
\end{bmatrix} =  \mathbf{M}\begin{bmatrix}
\overline{\z}^k \\
\overline{\g}^k \\
\x^k 
\end{bmatrix}+\begin{bmatrix}
\mathbf{0} \\
\bR_2\overline{\A^T}\y \\
\mu\bR_3\overline{\I}_{Mc}\bR_2\overline{\A^T}\y,
\end{bmatrix},
\label{EquMat}
\end{equation}
where $\mathbf{M}$ is

\noindent
\resizebox{\linewidth}{!}{\arraycolsep=2.5pt%
$
\begin{bmatrix}

   \I-\bR_1 & \mathbf{0} & \bR_1\overline{\A} \\
   \bR_2\overline{\A^T}\overline{\I}_{Nr}(\bR_1-\I) & \I-\bR_2 & -\bR_2\overline{\A^T}\overline{\I}_{Nr}\bR_1\overline{\A} \\
   \mu\bR_3\overline{\I}_{Mc}\bR_2\overline{\A^T}\overline{\I}_{Nr}(\bR_1-\I) &\mu\bR_3\overline{\I}_{Mc}(\I-\bR_2) &\I-\mu\bR_3\overline{\I}_{Mc}\bR_2\overline{\A^T}\overline{\I}_{Nr}\bR_1\overline{\A}
   \end{bmatrix}
$}


In the fixed point analysis, note that for any fixed point $\x^{\star}$, the random updates of $\z$ mean that we require that $\z_{I_i}^j=\A_{I_i}^{J_j}\x_J^\star$ for all $i$ and $j$, so that the fixed $\z^{\star}$ must be of the form $\z^{\star}=\overline{\A}\x^\star$. So we need:
\begin{equation}
\begin{bmatrix}
\overline{\A}\x^{\star} \\
\overline{\g}^{\star} \\
\x^{\star} 
\end{bmatrix} =  \mathbf{M}\begin{bmatrix}
\overline{\A}\x^{\star} \\
\overline{\g}^{\star} \\
\x^\star 
\end{bmatrix}+\begin{bmatrix}
\mathbf{0} \\
\bR_2\overline{\A^T}\y \\
\mu\bR_3\overline{\I}_{Mc}\bR_2\overline{\A^T}\y
\end{bmatrix}.
\label{EquMat2}
\end{equation}
Since the first line of Eq.\ref{EquMat2} is an identity, we only focus on the second and third line. The second line can be expressed as:
\begin{equation}
\bR_2  \left( \overline{\A^T}  (\y- \A\x^*) 		- \overline{\g}^*\right)=\mathbf{0}.
\end{equation}
As $\bR_2$ is a random diagonal matrix, this implies that $ \g^*=\overline{\A^T}  (\y- \A\x^*) $. The third line, after the deformation, can be expressed as:
\begin{equation}
    \bR_3\overline{\I}_{Mc}\bR_2\left( \overline{\A^T}\left( \y-\A\x^{\star}\right)-\overline{\g}^\star \right)+\bR_3\overline{\I}_{Mc}\overline{\g}^\star=\mathbf{0}
\end{equation}
which suggest that $\g^\star\equiv\overline{\I}_{Mc}\overline{\g}^\star\equiv2\A^T(\y-\A\x^\star)=\mathbf{0}$. This proves that the fixed point is the least square solution. Our empirical results show convergence to the fixed point. A theoretical analysis and formal convergence proof is in preparation.

\section{Conclusion}
BSGD can be viewed as an improvement of our previous CSGD algorithm. It mainly focus on the case where a distributed network is adopted to reconstruct a large scale CT image/volume in parallel and the nodes in the network have limited access to both projection data and volume. When noise is Gaussian type, which is a common case in CT reconstruction area, iteration results obtained by BSGD approaches closer to the least square solution than other mature CT reconstruction algorithms such as SIRT, CAV and our previously CSGD method. Compared with the other optimization algorithm proposed in machine learning area, such as SVRG, ORBCDVD, the BSGD has higher computation efficiency. It also has the ability to address the sparse view CT reconstruction by combining itself with TV regularization. Simulations prove that both BSGD and BSGD-TV have the ability to be applied on a distributed network.




\ifCLASSOPTIONcaptionsoff
  \newpage
\fi

\vfill


\end{document}